\documentstyle[preprint,aps]{revtex}
\begin{document}
\preprint{}
\draft
\title{Canonical quantization of spontaneously broken topologically massive
gauge theory}
\author{Lusheng Chen, Gerald Dunne, Kurt Haller, and Edwin Lim-Lombridas}
\address{Department of Physics, University of Connecticut, Storrs, Connecticut
06269 U.S.A.}
\date{\today}
\maketitle
\begin{abstract}
In this paper we investigate the canonical quantization of a non-Abelian
topologically massive Chern-Simons theory in which the gauge fields are
minimally coupled to a multiplet of scalar fields in such a way that the gauge
symmetry is spontaneously broken. Such a model produces the Chern-Simons-Higgs
mechanism in which the gauge excitations acquire mass both from the
Chern-Simons term and from the Higgs-Kibble effect. The symmetry breaking is
chosen to be only partially broken, in such a way that in the broken vacuum
there remains a residual non-Abelian symmetry. We develop the canonical
operator structure of this theory in the broken vacuum, with particular
emphasis on the particle-content of the fields involved in the
Chern-Simons-Higgs mechanism. We construct the Fock space and express the
dynamical generators in terms of creation and annihilation operator modes. The
canonical apparatus is used to obtain the propagators for this theory, and we
use the Poincar\'e generators to demonstrate the effect of Lorentz boosts on
the particle states.
\end{abstract}
\narrowtext

\section{Introduction}
Field theories in $(2+1)$-dimensional space-time exhibit many interesting and
important properties related to the masses of the particle excitations of the
quantum fields. For example, gauge theories involving a Chern-Simons term
support massive gauge field excitations \cite{J1,sdrj}, which differ from the
Higgs-Kibble excitations produced in conventional spontaneous symmetry breaking
\cite{Higgs}. The combination of {\it both\/} spontaneous symmetry breaking
{\it and\/} a Chern-Simons term for the gauge field leads to the
Chern-Simons-Higgs (CSH) mechanism, in which the physical fields are transmuted
in a process that combines the Chern-Simons and Higgs-Kibble mass-generating
effects in a particularly interesting and instructive manner.

An analysis of the covariant gauge field propagator \cite{PR1,hlousek}
indicates the presence of two distinct mass poles, with masses given as
complicated functions of the Higgs mass scale (set by the tree-approximation
minimum of the symmetry breaking potential) and the Chern-Simons mass scale
(coming from the Chern-Simons coupling parameter which has dimensions of mass
in three dimensional space-time). The two distinct mass poles may also be seen
in a factorization of the Chern-Simons-Proca equations of motion \cite{paul}. A
Schr\"odinger representation approach \cite{GD1} provides a simple physical
picture based on a quantum mechanical analogue which identifies the two masses
precisely with the two characteristic frequencies of the planar quantum
mechanical model of charged particles moving in both a uniform magnetic field
and a harmonic potential well. In this present paper we investigate field
theoretic aspects of the Chern-Simons-Higgs mechanism more deeply, presenting a
detailed analysis of the canonical quantization of spontaneously broken
Chern-Simons theories. In this work, we pay particular attention to the
relation between the quantized fields and their particle excitation modes and
to the structure of the Poincar{\'e} generators as functionals of these
particle excitation operators.

We have chosen to consider a non-Abelian theory in which the non-Abelian gauge
symmetry is spontaneously broken in a manner that preserves a residual
non-Abelian symmetry in the broken vacuum. This choice is motivated by the
question of how a spontaneously broken Chern-Simons theory `knows' to
quantum-mechanically protect the residual non-Abelian gauge symmetry from
topologically nontrivial gauge transformations. For non-Abelian Chern-Simons
theories, quantum consistency \cite{J1} requires that the Chern-Simons coupling
parameter takes quantized integer values, in appropriate units. Qualitatively,
this consistency condition is reminiscent of Dirac's quantum mechanical
quantization condition for the magnetic monopole \cite{PDirac}, but since the
Chern-Simons theory is a {\it field theory\/} further subtleties (such as
renormalization) arise. Pisarski and Rao \cite{PR} showed that for a
Chern-Simons-Yang-Mills theory (with no matter fields or symmetry breaking) a
consistent one-loop renormalization involves a finite additive renormalization
of the Chern-Simons mass, with the finite shift depending on the gauge group
and being such that the integer quantization condition is preserved. Subsequent
calculations have confirmed the conjecture \cite{PR} that there are no further
radiative corrections to this result \cite{G}. Perturbative analyses of Abelian
Chern-Simons theories subject to spontaneous symmetry breaking confirm the
topological basis of the integer quantization of the renormalization of the
Chern-Simons term \cite{khlebnikov1,spiridonov,khlebnikov2}. This work has
shown that, in the Abelian case, in which topological arguments do not apply,
the Chern-Simons mass receives a shift, in the broken vacuum, which is not an
integer, but a complicated function of the various bare mass scales\footnote{In
Ref.~\cite{KMP} it is suggested that this shift should not be interpreted as a
finite renormalization of the Chern-Simons mass, but rather as an indication of
the appearance of parity-violating terms in the effective action. This
reformulation of the result extends the Coleman-Hill theorem \cite{coleman},
concerning the absence of loop corrections to the Chern-Simons mass, to the
case of Abelian spontaneously broken Chern-Simons theories.}. In a
spontaneously broken non-Abelian Chern-Simons theory, with a {\it completely\/}
broken symmetry in which the invariance of the effective theory to gauge
transformations is no longer supported \cite{khlebnikov2}, similar behavior was
found\footnote{Note that the explicit formula for the finite shift reported
in Ref.~\cite{khlebnikov2} is incorrect, although this does not affect the
qualitative conclusions of that paper. The corrected integral appears in
Ref.~\cite{CDHL}.}.  More interesting is the situation in which the non-Abelian
gauge symmetry is only {\it partially\/} broken, leaving a residual {\it
non-Abelian\/} symmetry in the broken phase. The presence of the non-Abelian
residual symmetry suggests that the Chern-Simons coupling parameter should
again be quantized, and indeed a direct perturbative computation \cite{CDHL}
shows that the Chern-Simons coupling parameter receives a quantized finite
shift which preserves the quantum consistency condition in the broken
vacuum.\footnote{This model was first considered in Ref.~\cite{Khare}, but the
opposite conclusion was reported.}  This work confirms the validity of the
effective theory that describes the quantum fluctuations of the field about the
spontaneously broken vacuum; and it
motivates an investigation into the origins of the massive propagating particle
excitations of this model, and the mechanisms by which they obtain their mass.

In this paper, we consider the canonical quantization of such a non-Abelian
model, with a partially broken symmetry leaving a residual non-Abelian symmetry
in the broken phase, and develop the underlying  dynamical theory. We make
explicit the representation of the operator-valued fields in terms of
excitations that correspond to observable, propagating particles in the
spontaneously broken vacuum. We formulate the model in $(2+1)$-dimensional
Minkowski space-time and for definiteness we consider an octet of $SU(3)$ gauge
fields interacting with a triplet of scalar fields in the fundamental
representation of $SU(3)$. The scalar fields $\Phi$ are coupled
gauge-invariantly to the gauge fields, and self-coupled through a quartic
potential $V(\Phi^\dagger\Phi) = \mu^2\Phi^\dagger\Phi-\case 1/2
h(\Phi^\dagger\Phi)^2$, where $\mu^2>0$ and $h>0$, so that, in the tree
approximation, the scalar fields have nonvanishing vacuum expectation values.
The vacuum expectation values of the three constituent fields in $\Phi$ are
chosen so that the residual ``effective'' fields, which represent fluctuations
of these scalar fields about their tree approximation vacuum expectation
values, still maintain an unbroken $SU(2)$ symmetry in their coupling to the
gauge fields. In the canonical quantization of this model we construct
time-dependent fields in an interaction picture that includes, in the ``free''
Hamiltonian that drives it, the interaction terms that become bilinear in
fields when the charged scalar field $\Phi$ is expanded about its constant
vacuum expectation value. We use these time-dependent interaction-picture
fields to evaluate the propagators.  And finally, we construct the particle
states that correspond to the two different mass singularities in the
propagator for this model. We express the trilinear and quartic interaction
Lagrangian as a functional of these interaction-picture fields, and obtain a
set of vertices that can be used to describe the theory.  In addressing these
problems, we make use of technical developments that originated from separate
earlier work by the authors \cite{GD1,HLL1,HLL2,HLL3,KH1,GD2}.

In Section II, we formulate the model and describe the spontaneous symmetry
breaking process.  In Section III,  we construct the required Fock spaces,
express the scalar and gauge fields as
superpositions of particle and ghost excitations, and implement Gauss's law and
the gauge condition.  In Section IV, we construct the interaction-picture
scalar and gauge fields; and we evaluate their time-ordered vacuum
expectation values in the spontaneously broken vacuum state, to obtain the
propagators for this theory.  In Section V, we construct the Poincar\'e
generators for this theory, demonstrate the validity of the Poincar\'e algebra,
and evaluate the effect of Lorentz boosts on each of the massive gluon states.
Detailed forms of the interaction Lagrangian are given in an Appendix.

\section{Formulation of the model}
The Lagrangian for this model is given by\footnote{The implied summations over
repeated Latin superscripts, such as $a$, $b$, and $c$, are from 1 to 8 unless
otherwise specified.}
\begin{eqnarray}
{\cal L} &=& -\textstyle{\frac{1}{4}}{\cal F}_{\mu\nu}^{a}{\cal
F}^{a\mu\nu}+\textstyle{\frac{1}{4}}m\epsilon^{\mu\nu\rho}({\cal
F}_{\mu\nu}^{a}A_{\rho}^{a}+\textstyle{\frac{2}{3}}ef^{abc}A_{\mu}^{a}
A_{\nu}^{b}A_{\rho}^{c}) \nonumber \\
&+& (D^{\mu}\Phi)^{\dagger}(D_{\mu}\Phi) +\mu^{2}\Phi^{\dagger}\Phi-
\textstyle{\frac{1}{2}}h(\Phi^{\dagger}\Phi)^{2}
+ {\cal L}_{\text{fp}}\\
& + & \textstyle{\frac{1}{2}}(1-\gamma)G^{a}G^{a}-\left[\partial_{\mu}A^{a\mu}
-ie(1-\gamma)(\left\langle\Phi\right\rangle_{0}^{\dagger}
\lambda^{a}\Phi^{\prime}-
\Phi^{\prime \dagger}\lambda^{a}\left\langle\Phi\right\rangle_{0})\right]G^{a},
\label{eq:definelag}
\end{eqnarray}
where ${\cal F}_{\mu\nu}^{a}$ designates the $SU(3)$ gauge field strength
\begin{equation}
{\cal F}_{\mu\nu}^{a} = \partial_{\mu}A_{\nu}^{a}-\partial_{\nu}A_{\mu}^{a} -
2ef^{abc}A_{\mu}^{b}A_{\nu}^{c};
\label{eq:defnonabF}
\end{equation}
we denote by $F_{\mu\nu}^{a}$ the ``Abelian'' part of the field strength,
\begin{equation}
F_{\mu\nu}^{a}=\partial_{\mu}A_{\nu}^{a}-\partial_{\nu}A_{\mu}^{a},
\label{eq:defabF}
\end{equation}
and $f^{abc}$ represents the $SU(3)$ structure constants. The covariant
derivative of the scalar triplet $D_{\mu}\Phi$ is given by
\begin{equation}
D_{\mu}\Phi = \partial_{\mu}\Phi+ie\lambda^{a}A_{\mu}^{a}\Phi,
\label{eq:defcovder}
\end{equation}
where $\lambda^{a}$ represents the Gell-Mann matrices which satisfy the
commutation relations $[\lambda^a,\lambda^b]=2if^{abc}\lambda^c$. The
Lagrangian also contains the gauge-fixing term, with gauge-fixing parameter
$\gamma$, for the covariant gauge\,---\,in this case, the t'Hooft gauge, which
involves both the tree-approximation vacuum expectation value (v.e.v.) of the
scalar field $\left\langle\Phi\right\rangle_{0}$ and the fluctuation of the
scalar field about that vacuum expectation value $\Phi^{\prime} =
\Phi-\left\langle\Phi\right\rangle_{0}$. ${\cal L}_{\text{fp}}$ is the part of
the Lagrangian that couples the gauge fields to the Faddeev-Popov ghosts, and
is given by
\begin{equation}
{\cal L}_{\text{fp}} = i\partial_{\mu}\sigma_{\rm
f}^{a}\partial^{\mu}\sigma_{\rm p}^{a}+
2ief^{abc}A_{\mu}^{a}\sigma_{\rm f}^{b}\partial^{\mu}\sigma_{\rm p}^{c},
\label{eq:fadpopL}
\end{equation}
where $\sigma_{\rm f}^{a}$ and $\sigma_{\text{p}}^{\rm a}$ are the two
self-adjoint operator-valued anticommuting scalar Faddeev-Popov fields.

We choose a scheme for breaking the $SU(3)$ symmetry that preserves an
$SU(2)$ symmetry in the effective Lagrangian. In the tree-approximation vacuum
state for this effective Lagrangian, the self-interaction
$V(\Phi^{\dagger}\Phi)$ takes on its classical minimum value for the
tree-approximation vacuum expectation value
$\left\langle\Phi\right\rangle_{0}$. A choice for
$\left\langle\Phi\right\rangle_{0}$ that satisfies this requirement is
\begin{equation}
\left\langle\Phi\right\rangle_{0} = \frac{v}{\sqrt{2}}\left( \begin{array}{c}
0 \\ 0 \\ 1
\end{array}   \right) \equiv
\frac{v}{\sqrt{2}}\,\left\langle\phi\right\rangle_0,
\label{eq:vacexpphi}
\end{equation}
where $v=(2\mu^{2}/h)^{1/2}$. To analyze this model in the broken vacuum, we
expand the scalar field $\Phi$ in terms of its fluctuations about the v.e.v.
$\left\langle\Phi\right\rangle_{0}$
\begin{equation}
\Phi^\prime = \Phi - \left\langle\Phi\right\rangle_{0}
\end{equation}
and expand the Lagrangian as
\begin{equation}
{\cal L} = {\cal L}_0 + {\cal L}_1 + {\cal L}_2.
\label{eq:calLL}
\end{equation}
Here ${\cal L}_0$ represents the ``free'' Lagrangian, in which the interaction
have been shut off, and ${\cal L}_1$ and ${\cal L}_2$ represent terms that are
first and second order in $e$, respectively. Note that there are several
coupling constants and mass scales to consider when making this expansion, and
we need to be specific about how coupling constants are ``shut off'' in taking
${\cal L}$ to its noninteracting limit ${\cal L}_0$. The Chern-Simons coupling
constant $m$ has dimensions of mass, as do $e^2$ (the square of the
scalar-gauge coupling), $v^2$ (the square of the magnitude of the scalar field
v.e.v.), and $ev$. The noninteracting limit ${\cal L}_0$ of the full Lagrangian
${\cal L}$ is defined to be the limit $e\rightarrow0$ and $h\rightarrow0$ with
the ``Higgs'' mass scale $ev$ kept constant, and the Chern-Simons mass scale
unaffected. Then the noninteracting Lagrangian is
\begin{eqnarray}
{\cal L}_0 &=& -\frac{1}{4}\,F^a_{\mu\nu}F^{a\mu\nu} +
\frac{m}{4}\,\epsilon^{\mu\nu\rho}F^a_{\mu\nu}A^a_\rho +
\frac{e^2v^2}{4}\,A_\mu^aA^{b\mu}
\left\langle\phi\right\rangle_0^\dagger\{\lambda^a,\lambda^b\}
\left\langle\phi\right\rangle_0 +
\left|\partial_\mu\Phi^\prime\right|^2\nonumber\\
&-& \frac{\mu^2}{2}\left(\left\langle\phi\right\rangle_0^\dagger\Phi^\prime +
\Phi^{\prime\dagger}\left\langle\phi\right\rangle_0\right)^2 +
i\frac{ev}{\sqrt{2}}\,A_\mu^a
\left[(\partial^\mu\Phi^\prime)^\dagger\lambda^a\left\langle\phi\right\rangle_0
-\left\langle\phi\right\rangle_0^\dagger\lambda^a\partial^\mu\Phi^\prime\right]
\nonumber\\
&-& \left[\partial_\mu A^{a\mu} -
i\frac{ev}{\sqrt{2}}\,(1-\gamma)\left(\left\langle\phi
\right\rangle_0^\dagger\lambda^a\Phi^\prime - \Phi^{\prime\dagger}\lambda^a
\left\langle\phi\right\rangle_0\right)\right]\!G^a\nonumber\\
&+& \frac{1}{2}\,(1-\gamma)G^aG^a +
i\partial_\mu\sigma^a_{\text{f}}\partial^\mu\sigma^a_{\text{p}}.
\end{eqnarray}
The ${\cal O}(e)$ interaction Lagrangian is
\begin{eqnarray}
{\cal L}_1 &=& e\left\{f^{abc}F_{\mu\nu}^aA^{b\mu}A^{c\nu} -
\frac{m}{3}\,\epsilon^{\mu\nu\rho}f^{abc}A^a_\mu A^b_\nu A_\rho^c +
2if^{abc}A_\mu^a\sigma^b_{\text{f}}\partial^\mu\sigma^c_{\text{p}}
\right.\nonumber\\
&&\ \ \ \ \ +\ \frac{ev}{2\sqrt{2}}\,A_\mu^aA^{b\mu}
\left(\Phi^{\prime\dagger}\{\lambda^a,\lambda^b\}
\left\langle\phi\right\rangle_0 + \left\langle\phi\right\rangle_0^\dagger
\{\lambda^a,\lambda^b\}\Phi^\prime\right)\nonumber\\
&&\ \ \ \ \ -\
\left.iA_\mu^a\left[\Phi^{\prime\dagger}\lambda^a\partial^\mu\Phi^\prime -
\left(\partial^\mu\Phi^\prime\right)^\dagger\lambda^a\Phi^\prime\right] -
\sqrt{2}\,\frac{\mu^2}{ev}\,\left|\Phi^\prime\right|^2
\left(\left\langle\phi\right\rangle_0^\dagger\Phi^\prime +
\Phi^{\prime\dagger}\left\langle\phi\right\rangle_0\right)\right\}
\label{eq:calL1}
\end{eqnarray}
and the ${\cal O}(e^2)$ interaction Lagrangian is
\begin{equation}
{\cal L}_2 = e^2\left(-f^{abc}f^{ade}A_\mu^bA^{d\mu}A_\nu^cA^{e\nu} -
\frac{\mu^2}{e^2v^2}\left|\Phi^\prime\right|^4 +
\frac{1}{2}\,A_\mu^aA^{b\mu}\Phi^{\prime\dagger}
\{\lambda^a,\lambda^b\}\Phi^\prime\right).
\label{eq:calL2}
\end{equation}
We note that the presence of the Chern-Simons term in the original Lagrangian
Eq.~(\ref{eq:calLL}) introduces a new quadratic piece $\sim\epsilon FA$ in
${\cal L}_0$ and a new 3-gluon vertex piece $\sim\epsilon AAA$ in ${\cal L}_1$.

To identify the physical and unphysical fields in the broken vacuum, we first
express $\Phi^\prime$ in terms of {\it real\/} scalar fields
\begin{equation}
\Phi^{\prime} = \frac{1}{\sqrt{2}}\left( \begin{array}{rcl}
i\xi^{4} &+& \xi^{5} \\ i\xi^{6}&+&\xi^{7} \\ -i\xi^{8}&+&\psi
\end{array}  \right).
\label{eq:phiprime}
\end{equation}
Then, using the explicit form given in Eq.~(\ref{eq:vacexpphi}) of the v.e.v.
$\left\langle\phi\right\rangle_0$, together with the Gell-Mann matrix
anticommutation relations
\begin{equation}
\{\lambda^a,\lambda^b\} = \case 4/3 \delta^{ab}\,{\bf 1} + 2d^{abc}\lambda^c,
\end{equation}
we can write the free Lagrangian ${\cal L}_0$ as
\begin{eqnarray}
{\cal L}_0 &=& -\frac{1}{4}\,F_{\mu\nu}^aF^{a\mu\nu} +
\frac{m}{4}\,\epsilon^{\mu\nu\rho}F_{\mu\nu}^aA_\rho^a +
\frac{1}{2}\sum_{a=4}^8M_{(a)}^2A_\mu^aA^{a\mu} +
\sum_{a=4}^8M_{(a)}A^{a\mu}\partial_\mu\xi^a\nonumber\\
&+&\frac{1}{2}\sum_{a=4}^8\partial_\mu\xi^a\partial^\mu\xi^a +
\frac{1}{2}\,\partial_\mu\psi\partial^\mu\psi - \mu^2\psi^2 +
\frac{1}{2}\,(1-\gamma)G^aG^a\nonumber\\
&-& \partial_\mu A^{a\mu}G^a + (1-\gamma)\sum_{a=4}^8M_{(a)}\xi^aG^a +
i\partial_\mu\sigma^a_{\text{f}}\partial^\mu\sigma^a_{\text{p}}.
\label{eq:calL0}
\end{eqnarray}
Here the symmetry breaking mass scales $M_{(a)}$ are given by
\begin{equation}
M_{(a)}=\left\{ \begin{array}{rcll}
M_{\rm D}&=&ev & a=4,5,6,7 \\
M_{\rm S}&=&\displaystyle\frac{2}{\sqrt{3}}\,ev\ \ \  & a=8
\end{array} \right..
\label{eq:mass}
\end{equation}
{}From Eq.~(\ref{eq:calL0}), we recognize $\psi$ as the Higgs scalar field,
with mass $\sqrt{2}\,|\mu|$, and $\xi^a$ $(a=4,\ldots,8)$ as massless
unphysical scalar fields. Furthermore, the gauge fields $A_\mu^a$ $(a=1,2,3)$
have a quadratic Lagrangian of the Maxwell-Chern-Simons form, while the gauge
fields $A_\mu^a$ $(a=4,\ldots,8)$ have an additional Proca-like quadratic term
with mass scale parameters $M_{(a)}$ as given in Eq.~(\ref{eq:mass}).

The interaction Lagrangians ${\cal L}_1$ and ${\cal L}_2$ can also be expanded
in terms of the real fields in Eq.~(\ref{eq:phiprime}) and the symmetry
breaking mass scales in Eq.~(\ref{eq:mass}), and the resulting expansions are
recorded in Appendix A. It is important to observe that (as expected) the gauge
field $A_\mu^a$ $(a=1,2,3)$ form an $SU(2)$ triplet corresponding to the
residual $SU(2)$ symmetry of the broken vacuum. It proves convenient to group
the real scalar fields into $SU(2)$ ``isospinors'':
\begin{equation}
\Psi_{1} = \frac{1}{\sqrt{2}}\left( \begin{array}{rcl}
i\xi^{4}&+&\xi^{5} \\ i\xi^{6}&+& \xi^{7}
\end{array}  \right),
\label{eq:psi1cap}
\end{equation}
\begin{equation}
\Psi_{2} = \frac{1}{\sqrt{2}}\left( \begin{array}{rcl}
i\xi^{4}&+&\xi^{5} \\ \psi&-&i\xi^{8}
\end{array}  \right),
\end{equation}
and
\begin{equation}
\Psi_{3} = \frac{1}{\sqrt{2}}\left(\begin{array}{rcl}
i\xi^{6}&+&\xi^{7} \\ \psi&-&i\xi^{8}
\end{array}  \right).
\label{eq:psi3cap}
\end{equation}
With this notation, the fields $A_\mu^a$ $(a=1,2,3)$ couple to $\Psi_1$ in an
$SU(2)$-invariant manner, while the two gauge field doublets
$(A_\mu^4,A_\mu^5)$ and $(A_\mu^6,A_\mu^7)$ couple to $\Psi_2$ and $\Psi_3$ so
that the part of the isospin invariance that corresponds to rotation in the
$i=1,2$ plane is preserved; but this latter interaction is not invariant to
rotation in the entire isospin space. The remaining gauge field $A_\mu^8$ is an
$SU(2)$ singlet.

In earlier work on Abelian theories with Chern-Simons interactions
\cite{HLL1,HLL2,HLL3}, we implemented Gauss's law and developed a canonical
formulation for the entire Lagrangian, with all interactions included.  In a
non-Abelian gauge theory, such a program becomes much more problematical.  We
will therefore implement Gauss's law only for the partial theory described by
${\cal L}_{0}$.  In this case, however, because of the spontaneously broken
symmetry, even the Abelian ${\cal L}_{0}$ contains part of the
interaction\,---\,not only the part of the $\Phi^4$ self-interaction implicit
in the spontaneously broken vacuum state, but also the part of the
gauge-invariant coupling of the gauge field to the ``charged'' scalar $\Phi$
that remains bilinear in operator-valued fields after $\Phi$ has been expressed
as $\Phi= \Phi^{\prime} + \left\langle\Phi\right\rangle_{0}$.  Although this
part of the interaction term is proportional to $e$, it does not vanish in the
``interaction-free'' limit, because $e$ combines with $h^{-1/2}$ to become one
of the masses that are kept constant in the ${\cal L}\rightarrow{\cal L}_{0}$
limit.  Implementing Gauss's law and the gauge condition, and developing the
canonical formulation of the part of the theory described by ${\cal L}_{0}$,
will enable us to construct the Fock space for the particle states observed in
the broken vacuum. In the course of this work, we will demonstrate the process
by which the masses that stem from the Higgs-Kibble effect \cite{Higgs} combine
with the topological mass to form the masses of the propagating modes of the
gauge field in this model.  ${\cal L}_{0}$, defined as we have specified here,
is the Lagrangian that drives the interaction-picture fields when a
Higgs-Kibble effect occurs. The corresponding ``free'' Hamiltonian $H_0$, which
is the $e\rightarrow0$ limit of $H$ obtained by this same limiting process,
accounts for the particle spectrum of this model. Once ${\cal L}_{0}$ and $H_0$
have been identified, and Gauss's law and the covariant gauge condition have
been imposed, the resulting apparatus can be used to develop a Fock space as
well as propagators and vertices for evaluating the $S$-matrix elements and
renormalization constants for the full theory, with ${\cal L}_{1}$ and ${\cal
L}_{2}$ included.

The Euler-Lagrange equations determined by ${\cal L}_{0}$ are
\begin{equation}
\partial_{\mu}F^{a\nu\mu}-\textstyle{\frac{1}{2}}m
\epsilon^{\mu\rho\nu}F_{\mu\rho}^{a}-\partial^{\nu}G^{a}=
M_{(a)}^{2}A^{a\nu}+\partial^{\nu}\alpha^{a},
\label{eq:gaugefield}
\end{equation}
\begin{equation}
\partial_{\mu}A^{a\mu}-(1-\gamma)\alpha^{a}  =  (1-\gamma)G^{a},
\label{eq:gaugecondition}
\end{equation}
\begin{equation}
\partial_{\mu}\partial^{\mu}\psi+2\,\mu^{2}\psi  =  0,
\end{equation}
and
\begin{equation}
\partial_{\mu}\partial^{\mu}\xi^{a}+M_{(a)}\partial_{\mu}A^{a\mu}  =
(1-\gamma)M_{(a)}G^{a}, \label{eq:goldstonfield}
\end{equation}
where $\alpha^{a}=M_{(a)}\xi^{a},$
and
\begin{equation}
\partial_{\mu}\partial^{\mu}\sigma_{\text{f}}^a =
\partial_{\mu}\partial^{\mu}\sigma_{\text{p}}^a  =  0.
\end{equation}
{}From these equations, we get
\begin{equation}
\partial_{\mu}\partial^{\mu}G^{a}=-(1-\gamma)M_{(a)}^{2}G^{a}.
\label{eq:ghostfield}
\end{equation}
Equation~(\ref{eq:gaugefield}) represents the Maxwell-Ampere law (for
${\nu}=1,2$) as well as Gauss's law (for ${\nu}=0$); however, as is to be
expected in covariant gauges, this equation differs from the classical form of
Maxwell-Ampere and Gauss's laws by the gauge-fixing term\,---\,in this case,
$\partial^{\nu}G^{a}+\partial^{\nu}\alpha^{a}$.  Implementation of the correct
form of these laws will have the effect of defining a subspace for the
dynamical time-evolution of state vectors in which the gauge-fixing term will
have vanishing matrix elements. Equation~(\ref{eq:gaugecondition}) will be used
to impose the covariant gauge condition: $\gamma =0$ corresponds to the
Feynman, and $\gamma =1$ to the Landau version of the covariant (t'Hooft)
gauge.

To quantize this theory, we need to express the Hamiltonian in terms
of the canonical momenta given by $\Pi^{a\mu}={\partial{\cal
L}_{0}}/{\partial(\partial_{0}A_{\mu}^{a})}$. These canonical momenta are:
\begin{equation}
\Pi^{a\mu} = F^{a\mu 0} +
\textstyle{\frac{1}{2}}m\epsilon^{0\mu\nu}A_{\nu}^{a}-
g^{0\mu}G^{a},
\label{eq:gaugemoment}
\end{equation}
\begin{equation}
\Pi_{\psi} = \partial_{0}\psi,
\end{equation}
\begin{equation}
\Pi_{\xi}^{a} = \partial_{0}\xi^{a}+M_{(a)}A_{0}^{a},
\end{equation}
\begin{equation}
\Pi_{\rm f}^{a}= i\partial_{0}\sigma_{\rm p}^{a},
\end{equation}
and
\begin{equation}
\Pi_{\rm p}^{a} = -i\partial_{0}\sigma_{\rm f}^{a};
\end{equation}
$\Pi_{\rm f}^{a}$ and $\Pi_{\rm p}^{a}$ are the conjugate momenta to the fields
$\sigma_{\rm f}^{a}$ and $\sigma_{\rm p}^{a}$, respectively.

The only equation that does not contain any time-derivatives of fields (and
therefore is a constraint) is $\Pi^{a0} = -G^a$. This constraint is manifestly
consistent with canonical (Poisson) equal-time commutation rules, which we
impose. The equal-time commutation rules (ETCR) are:
\begin{equation}
[A_{l}^{a}({\bf x}),\Pi_{n}^{b}({\bf y})] =
i\delta_{ln}\,\delta^{ab}\,\delta({\bf
x}-{\bf y}),
\label{eq:commutator1}
\end{equation}
\begin{equation}
[A_{0}^{a}({\bf x}),G^{b}({\bf y})]  =  -i\delta^{ab}\,\delta({\bf x}-{\bf y}),
\label{eq:commutator2}
\end{equation}
\begin{equation}
[\xi^{a}({\bf x}),\Pi_{\xi}^{b}({\bf y})] =  i\delta^{ab}\,\delta({\bf x}-{\bf
y}),
\label{eq:commutator3}
\end{equation}
\begin{equation}
[\psi({\bf x}),\Pi_{\psi}({\bf y})]  = i\delta({\bf x}-{\bf y}),
\label{eq:commutator4}
\end{equation}
and all other commutators among these fields are zero. The anticommutation
rules for the Faddeev-Popov ghost fields are
\begin{equation}
\{\sigma_{\text{f}}^a({\bf x}),\Pi_{\text{f}}^b({\bf y})\} =
i\delta^{ab}\,\delta({\bf x-y}),
\end{equation}
\begin{equation}
\{\sigma_{\text{p}}^a({\bf x}),\Pi_{\text{p}}^b({\bf y})\} =
i\delta^{ab}\,\delta({\bf x-y}),
\end{equation}
and all other combinations anticommute.

The Hamiltonian density ${\cal H}_{0}$, determined by ${\cal L}_{0}$ and by
the canonical momenta, will be expressed as
\begin{equation}
{\cal H}_{0}=\sum_{a=1}^{8}{\cal H}^{a}+{\cal H}_{\psi} + {\cal H}_{\text{fp}};
\label{eq:Ham0}
\end{equation}
for $a=1,2,3$:
\begin{eqnarray}
{\cal H}^{a} & = &
\textstyle{\frac{1}{2}}\Pi_{l}^{a}\Pi_{l}^{a}+
\textstyle{\frac{1}{4}}F_{ln}^{a}F_{ln}^{a} +
\textstyle{\frac{1}{8}}m^{2}A_{n}^{a}A_{n}^{a} +
\textstyle{\frac{1}{2}}m\epsilon_{ln}A_{l}^{a}
\Pi_{n}^{a}\nonumber\\
&+&A_{0}^{a}(\partial_{l}\Pi_{l}^{a}-\textstyle{\frac{1}{4}}
m\epsilon_{ln}F_{ln}^{a})+G^{a}\partial_{l}A_{l}^{a}-
\textstyle{\frac{1}{2}}(1-\gamma)G^{a}G^{a};
\label{eq:Ha3}
\end{eqnarray}
for $a=4,5,6,7,8$:
\begin{eqnarray}
{\cal H}^{a} & = &
\textstyle{\frac{1}{2}}\Pi_{l}^{a}\Pi_{l}^{a}+
\textstyle{\frac{1}{4}}F_{ln}^{a}F_{ln}^{a}+
\textstyle{\frac{1}{2}}[\textstyle{\frac{1}{4}}m^{2}+
M_{(a)}^{2}]A_{n}^{a}A_{n}^{a} +
\textstyle{\frac{1}{2}}m\epsilon_{ln}A_{l}^{a}\Pi_{n}^{a} \nonumber \\
&+& A_{0}^{a}[\partial_{l}\Pi_{l}^{a}-\textstyle{\frac{1}{4}}m
\epsilon_{ln}F_{ln}^{a}-M_{(a)}\Pi_{\xi}^{a}]+
G^{a}[\partial_{l}A_{l}^{a}-(1-\gamma)M_{(a)}\xi^{a}]\nonumber\\
&-&M_{(a)}A_{l}^{a}\partial_{l}\xi^{a} -
\textstyle{\frac{1}{2}}(1-\gamma)G^{a}G^{a}+
\textstyle{\frac{1}{2}}\Pi_{\xi}^{a}\Pi_{\xi}^{a}+
\textstyle{\frac{1}{2}}\partial_{l}\xi^{a}\partial_{l}\xi^{a};
\label{eq:Ha8}
\end{eqnarray}
The other parts of ${\cal H}_0$ are
\begin{equation}
{\cal H}_{\psi} =
\textstyle{\frac{1}{2}}\Pi_{\psi}\Pi_{\psi}+
\textstyle{\frac{1}{2}}\partial_{l}\psi\partial_{l}\psi+\mu^{2}\psi^{2};
\end{equation}
and
\begin{equation}
{\cal H}_{\text{fp}} = i\Pi_{\text{p}}^a\Pi_{\text{f}}^a +
i\partial_{j}\sigma_{\text{f}}^{a}\partial_{j}\sigma_{\text{p}}^{a}.
\end{equation}
The Hamiltonian, $H_{0}=\int d{\bf x}\ {\cal H}_{0}({\bf x})$, is the ``free''
kinetic energy limit of the entire Hamiltonian, with the proviso that in this
model the free kinetic energy limit includes the part of the interaction term
in which the constant tree-approximation vacuum expectation value of
$\Phi$ combines with the charge $e$ to form a new constant, dimensionally a
mass, whose operator-valued coefficient is bilinear in fields.  This part of
the interaction is not shut off in the $H\rightarrow H_{0}$ limit, and is
absorbed into a generalized, more encompassing kinetic energy operator $H_{0}$.

\section{Particle states and Gauss's law}
Equation~(\ref{eq:gaugefield}), when $\nu =0$ and the canonical momenta replace
the time derivatives of fields, has the form\footnote{For notational
simplicity, we will, from here on, generally use a noncovariant notation in
which the subscript $l$ denotes a covariant component of a covariant quantity
(like $\partial_l$), a contravariant component of a contravariant quantity
(like $A_l$), or the contravariant component of the second rank tensor
$\Pi_l$.}
\begin{equation}
\partial_{l}\Pi_{l}^{a}+\textstyle{\frac{1}{2}}m
\epsilon_{ln}\partial_{l}A_{n}^{a}-M_{(a)}\Pi_{\xi}^{a}=\partial^{0}G^{a}.
\label{eq:Gauss's}
\end{equation}
The right-hand side of Eq.~(\ref{eq:Gauss's}) would have to vanish to express
Gauss's law.  But since $\partial^{0}G^{a}=0$ is not one of the Euler-Lagrange
equations, we therefore have to take some further measures to implement Gauss's
law.  For later reference, we will define the ``Gauss's law operator'' ${\cal
G}^a$ as
\begin{equation}
{\cal G}^a=\partial_{l}\Pi_{l}^{a}+\textstyle{\frac{1}{2}}m
\epsilon_{ln}\partial_{l}A_{n}^{a}-M_{(a)}\Pi_{\xi}^{a}.
\label{eq:Gscript}
\end{equation}

In order to describe the particle states of this theory, we must
construct a ``suitable'' representation for the  operator-valued fields in
terms of creation and annihilation operators  for the observable propagating
particles described by this model.  We expect these observable particle modes
to consist of Higgs scalars as well as  gauge field excitations\,---\,massive
gluons with both topological mass and  mass from the spontaneously broken
symmetry. The eight gauge fields in this model fall into three classes, and
should give rise to particle excitations with different mass: one $SU(2)$
triplet of gauge fields with $M_{(a)}=0$ and excitation modes that have only
topological mass; two doublets of gauge fields with $M_{(a)}=M_D$ excitations
whose mass depends on both the topological mass and $M_{\rm D}$; and a singlet
similar to the doublet, but with $M_{\rm D}$ replaced by  $M_{\rm S}$.  The
pole structure of the propagator \cite{PR1,hlousek} and earlier work on related
systems \cite{paul,GD1}
suggest that the gauge fields in the doublet and singlet sectors each have two
different massive gluon states.  The gauge fields in the  unbroken $SU(2)$
triplet have just a single gluon excitation mode. We will make an initial {\it
ansatz\/} that incorporates this set of particle states into the representation
of the gauge fields. If more particle states are needed than the ones included
in our {\it ansatz,\/} or if an entirely different set is required, it will be
impossible to construct a suitable representation using these excitation modes.
If fewer particle modes are sufficient, then it will be become evident that a
mode is redundant.  Mistakes in the tentative choices of particle modes will
therefore be self-correcting. Conversely, a consistent and suitable
representation of the gauge fields will confirm that the identification of the
particle excitations is correct.

The first requirement for a suitable representation is that it must be
consistent with the equal-time commutation rules given in
Eqs.~(\ref{eq:commutator1})--(\ref{eq:commutator4}). But it is apparent that
the  observable, propagating gluon modes listed above will not suffice to
represent all the commutation rules included in Eqs.~(\ref{eq:commutator1}) and
(\ref{eq:commutator2}).  Further modes, in the form of ghost excitations, are
required. These ghost modes are identical to the ones that appear in Abelian
Maxwell-Chern-Simons theory \cite{HLL1,HLL2,HLL3}, and that are also required
in $(3+1)$-dimensional QED (QED$_4$) in covariant and axial (except for the
spatial axial) gauges \cite{el3}. The excitation operators for the massive
gluons are the annihilation operator $a^c({\bf k})$ and its adjoint creation
operator $a^{c\dagger}({\bf k})$, which obey the commutation rule  $[a^c({\bf
k}),a^{d\dagger}({\bf q})]=\delta^{cd}\,\delta_{{\bf kq}}$. For the gauge
fields in the doublet and singlet sectors, the second
observable, propagating mode will be designated by the annihilation operator
$b^c({\bf k})$ and its adjoint creation operator $b^{c\dagger}({\bf k})$, which
obey the commutation rule $[b^c({\bf k}),b^{d\dagger}({\bf q})] =
\delta^{cd}\,\delta_{{\bf kq}}$.

Ghost excitation operators exist in pairs. In this work, we will use the ghost
annihilation operators  $a_Q^c({\bf k})$ and $a_R^c({\bf k})$ and their
respective adjoint creation operators $a_Q^{c\,\mbox{\normalsize $\star$}}({\bf
k})$ and $a_R^{c\,\mbox{\normalsize $\star$}}({\bf k})$ in the representations
of the gauge field. Ghost states have zero norm, but the single-particle ghost
states $a_Q^{c\,\mbox{\normalsize $\star$}}({\bf k})|0\rangle$ and
$a_R^{c\,\mbox{\normalsize $\star$}}({\bf k})|0\rangle$ have a nonvanishing
inner product; similar nonvanishing inner products also arise for $n$-particle
states with equal numbers of $Q$ and $R$ ghosts. These properties of the ghost
states are implemented by the commutator algebra
\begin{equation}
[a_Q^c({\bf k}),a_R^{d\,\mbox{\normalsize $\star$}}({\bf q})] = [a_R^c({\bf
k}),a_Q^{d\,\mbox{\normalsize $\star$}}({\bf q})] =\delta^{cd}\,\delta_{\bf kq}
\end{equation}
and
\begin{equation}
[a_Q^c({\bf k}),a_Q^{d\,\mbox{\normalsize $\star$}}({\bf q})] = [a_R^c({\bf
k}),a_R^{d\,\mbox{\normalsize $\star$}}({\bf q})] = 0,
\end{equation}
which, in turn, imply that the unit operator in the one-particle ghost (OPG)
sector is
\begin{equation}
1_{\text{OPG}} = \sum_{{\bf k}}\left[a_Q^{c\,\mbox{\normalsize $\star$}}({\bf
k})|0\rangle\langle
0|a_R^c({\bf k}) +a_R^{c\,\mbox{\normalsize $\star$}}({\bf k})|0\rangle\langle
0|a_Q^c({\bf k})\right];
\label{eq:1opg}
\end{equation}
the obvious generalization of Eq.~(\ref{eq:1opg}) applies in the $n$-particle
sectors.
The ghost excitations enable us to satisfy the equal-time commutation
relations, Eqs.~(\ref{eq:commutator1}) and (\ref{eq:commutator2}).

Another requirement we will impose on a ``suitable'' representation is that
the Gauss's law operator ${\cal G}^c({\bf x})$ be restricted to a linear
combination of ghost operators for a single kind of ghost.  We will stipulate
the specific requirement that  ${\cal G}^c({\bf x})$ be a superposition of
$a_Q^{c}({\bf k})$ and $a_Q^{c\,\mbox{\normalsize $\star$}}({\bf k})$
operators.  There is yet another criterion that a representation must satisfy
in order to be suitable: The gluon modes (propagating and ghost) must appear in
the Hamiltonian $H_0$  in such a manner that dynamical
time-evolution\,---\,i.e. translation by the time-displacement operator
$\exp\left(-iH_{0}t\right)$\,---\,never propagates state vectors into the
``dangerous'' part of Hilbert space in which inner products between the two
different types of ghost states drain  probability from observable particle
states.

We have found the required suitable representation of the fields by a
combination of unitary transformations similar to the ones used in previous
work \cite{HLL1,HLL2,HLL3} and of ``trial fields'' with arbitrary parameters
which we then adjusted to arrive at ``suitable'' field representations. For
example,
we used the trial field
\begin{eqnarray}
A_l^c({\bf x}) &=& \sum_{\bf k}[\alpha_1(k)\epsilon_{ln}k_n +
\alpha_2(k)k_l]\left[a^{c}({\bf
k})e^{i{\bf k \cdot x}}+ a^{c\dagger}({\bf k})e^{-i{\bf k \cdot
x}}\right]\nonumber\\
&+& \sum_{\bf k}i[\alpha_3(k)\epsilon_{ln}k_n + \alpha_4(k)k_l]\left[a^{c}({\bf
k})e^{i{\bf k \cdot x}}- a^{c\dagger}({\bf k})e^{-i{\bf k \cdot
x}}\right]\nonumber\\
&+&\sum_{\bf k}[\alpha_5(k)\epsilon_{ln}k_n + \alpha_6(k)k_l]\left[b^c({\bf
k})e^{i{\bf k \cdot x}}+b^{c\dagger}({\bf k})e^{-i{\bf k \cdot
x}}\right]\nonumber\\
&+& \sum_{\bf k}i[\alpha_7(k)\epsilon_{ln}k_n + \alpha_8(k)k_l]\left[b^c({\bf
k})e^{i{\bf k \cdot x}}- b^{c\dagger}({\bf k})e^{-i{\bf k \cdot
x}}\right]\nonumber\\
&+&\sum_{\bf k}[\alpha_9(k)\epsilon_{ln}k_n +
\alpha_{10}(k)k_l]\left[a_Q^c({\bf k})e^{i{\bf k \cdot
x}}+a_Q^{c\,\mbox{\normalsize $\star$}}({\bf k})e^{-i{\bf k \cdot
x}}\right]\nonumber\\
&+& \sum_{\bf k}i[\alpha_{11}(k)\epsilon_{ln}k_n +
\alpha_{12}(k)k_l]\left[a_Q^c({\bf k})e^{i{\bf k \cdot
x}}-a_Q^{c\,\mbox{\normalsize $\star$}}({\bf k})e^{-i{\bf k \cdot
x}}\right]\nonumber\\
&+&\sum_{\bf k}[\alpha_{13}(k)\epsilon_{ln}k_n +
\alpha_{14}(k)k_l]\left[a_R^c({\bf k})e^{i{\bf k \cdot
x}}+a_R^{c\,\mbox{\normalsize $\star$}}({\bf k})e^{-i{\bf k \cdot
x}}\right]\nonumber\\
&+& \sum_{\bf k}i[\alpha_{15}(k)\epsilon_{ln}k_n +
\alpha_{16}(k)k_l]\left[a_R^c({\bf k})e^{i{\bf k \cdot
x}}-a_R^{c\,\mbox{\normalsize $\star$}}({\bf k})e^{-i{\bf k \cdot x}}\right],
\label{eq:genA}
\end{eqnarray}
where $\alpha_1(k),\ldots,\alpha_{16}(k)$ are arbitrary real parameters.
Similar substitutions were made for the other fields in the model.  The
requirements of ``suitability'' were then translated into a set of equations
which was solved using a customized operator algebra manipulation package in
MATHEMATICA \cite{math}.  The resulting gauge field representations for the
$SU(2)$-symmetric triplet ($c=1,2,3$) that has topological mass only are
\begin{eqnarray}
A_l^{c}({\bf x}) &=& \sum_{\bf
k}\frac{8ik\epsilon_{ln}k_n}{m^{5/2}}\left[a_Q^c({\bf k})e^{i{\bf k\cdot
x}}-a_Q^{c\,\mbox{\normalsize $\star$}}({\bf k})e^{-i{\bf k\cdot
x}}\right]\nonumber\\
&+& (1-\gamma)\sum_{\bf k}\frac{2k_l}{m^{3/2}}\left[a_Q^c({\bf k})e^{i{\bf
k\cdot x}}+a_Q^{c\,\mbox{\normalsize $\star$}}({\bf k})e^{-i{\bf k\cdot
x}}\right]\nonumber\\
&-&\sum_{\bf k}\frac{4k^2k_l}{m^{7/2}}\left[a_Q^c({\bf k})e^{i{\bf k\cdot
x}}+a_Q^{c\,\mbox{\normalsize $\star$}}({\bf k})e^{-i{\bf k\cdot
x}}\right]\nonumber\\
&+&\sum_{\bf k}\frac{m^{3/2}k_l}{16k^3}\left[a_R^c({\bf k})e^{i{\bf k\cdot
x}}+a_R^{c\,\mbox{\normalsize $\star$}}({\bf k})e^{-i{\bf k\cdot
x}}\right]\nonumber\\
&-&\sum_{\bf k}\frac{\sqrt{\omega(k)}k_l}{\sqrt{2}mk}\left[a^c({\bf k})e^{i{\bf
k\cdot x}}+a^{c\dagger}({\bf k})e^{-i{\bf k\cdot x}}\right]\nonumber\\
&+&\sum_{\bf k}\frac{i\epsilon_{ln}k_n}{k\sqrt{2\omega(k)}}\left[a^c({\bf
k})e^{i{\bf k\cdot x}}-a^{c\dagger}({\bf k})e^{-i{\bf k\cdot x}}\right],
\label{eq:Ail}
\end{eqnarray}
\begin{eqnarray}
\Pi_l^{c}({\bf x}) &=& -\sum_{\bf k}\frac{4ikk_l}{m^{3/2}}\left[a_Q^c({\bf
k})e^{i{\bf k\cdot x}}-a_Q^{c\,\mbox{\normalsize $\star$}}({\bf k})e^{-i{\bf
k\cdot x}}\right]\nonumber\\
&+&(1-\gamma)\sum_{\bf k}\frac{\epsilon_{ln}k_n}{\sqrt{m}}\left[a_Q^c({\bf
k})e^{i{\bf k\cdot x}}+a_Q^{c\,\mbox{\normalsize $\star$}}({\bf k})e^{-i{\bf
k\cdot x}}\right]\nonumber\\
&+&\sum_{\bf k}\frac{6k^2\epsilon_{ln}k_n}{m^{5/2}}\left[a_Q^c({\bf k})e^{i{\bf
k\cdot x}}+a_Q^{c\,\mbox{\normalsize $\star$}}({\bf k})e^{-i{\bf k\cdot
x}}\right]\nonumber\\
&+&\sum_{\bf k}\frac{m^{5/2}\epsilon_{ln}k_n}{32k^3}\left[a_R^c({\bf
k})e^{i{\bf k\cdot x}}+a_R^{c\,\mbox{\normalsize $\star$}}({\bf k})e^{-i{\bf
k\cdot x}}\right]\nonumber\\
&+&\sum_{\bf k}\frac{imk_l}{2^{3/2}k\sqrt{\omega(k)}}\left[a^c({\bf k})e^{i{\bf
k\cdot x}}-a^{c\dagger}({\bf k})e^{-i{\bf k\cdot x}}\right]\nonumber\\
&+&\sum_{\bf k}\frac{\sqrt{\omega(k)}\epsilon_{ln}k_n}{2^{3/2}k}\left[a^c({\bf
k})e^{i{\bf k\cdot x}}+a^{c\dagger}({\bf k})e^{-i{\bf k\cdot x}}\right],
\label{eq:Pilmomentum}
\end{eqnarray}
\begin{eqnarray}
A_0^{c}({\bf x}) &=& -\sum_{\bf k}\frac{4k^3}{m^{7/2}}\left[a_Q^c({\bf
k})e^{i{\bf k\cdot x}}+a_Q^{c\,\mbox{\normalsize $\star$}}({\bf k})e^{-i{\bf
k\cdot x}}\right]\nonumber\\
&-&(1-\gamma)\sum_{\bf k}\frac{2k}{m^{3/2}}\left[a_Q^c({\bf k})e^{i{\bf k\cdot
x}}+a_Q^{c\,\mbox{\normalsize $\star$}}({\bf k})e^{-i{\bf k\cdot
x}}\right]\nonumber\\
&+&\sum_{\bf k}\frac{m^{3/2}}{16k^2}\left[a_R^c({\bf k})e^{i{\bf k\cdot
x}}+a_R^{c\,\mbox{\normalsize $\star$}}({\bf k})e^{-i{\bf k\cdot
x}}\right]\nonumber\\
&-&\sum_{\bf k}\frac{k}{m\sqrt{2\omega(k)}}\left[a^c({\bf k})e^{i{\bf k\cdot
x}}+a^{c\dagger}({\bf k})e^{-i{\bf k\cdot x}}\right],
\end{eqnarray}
and
\begin{eqnarray}
G^{c}({\bf x}) = \sum_{\bf k}\frac{8ik^2}{m^{3/2}}\left[a_Q^c({\bf k})e^{i{\bf
k\cdot x}}-a_Q^{c\,\mbox{\normalsize $\star$}}({\bf k})e^{-i{\bf k\cdot
x}}\right],
\label{eq:G}
\end{eqnarray}
where $\omega(k) = \displaystyle\sqrt{m^2 + k^2}$; and for the doublet and
singlet sectors with combined topological and ``Higgs-Kibble'' mass
($c=4,\ldots,8$), the fields are represented by
\begin{eqnarray}
A_{l}^{c}({\bf x}) &=& -\sum_{\bf k} \sqrt{\frac{\omega_c(k)}{2{\sf m}_c({\sf
m}_c+ \bar{\sf m}_c)}}\,\frac{k_l}{k}\left[a^{c}({\bf k})
e^{i{\bf k}\cdot{\bf x}}+a^{c\dagger}({\bf k})e^{-i{\bf k}\cdot{\bf
x}}\right]\nonumber\\
&+& \sum_{\bf k}\sqrt{\frac{{\sf m}_c}{2\omega_c(k)({\sf m}_c+\bar{\sf
m}_c)}}\,\frac{i\epsilon_{ln}k_n}{k}\left[a^{c}({\bf k})e^{i{\bf k}\cdot{\bf
x}}-a^{c\dagger}({\bf k})e^{-i{\bf k}\cdot{\bf x}}\right]\nonumber\\
&+&\sum_{\bf k}\sqrt{\frac{\bar{\omega}_c(k)}{2\bar{\sf m}_c({\sf m}_c+\bar{\sf
m}_c)}}\, \frac{ik_l}{k}\left[b^{c}({\bf k})e^{i{\bf k}\cdot{\bf
x}}-b^{c\dagger}({\bf k})e^{-i{\bf k}\cdot{\bf x}}\right] \nonumber\\
&-&\sum_{\bf k}\sqrt{\frac{\bar{\sf m}_c}{2\bar{\omega}_c(k)({\sf m}_c+\bar{\sf
m}_c)}}\, \frac{\epsilon_{ln}k_n}{k}\left[b^{c}({\bf k})e^{i{\bf k}\cdot{\bf
x}}+b^{c\dagger}({\bf k})e^{-i{\bf k}\cdot{\bf x}}\right] \nonumber\\
&-& \sum_{\bf k} \frac{4k^3k_l}{\kappa_{c}(\gamma){\sf m}_c\bar{\sf m}_c({\sf
m}_c-\bar{\sf m}_c)^{3/2}} \left[a_{Q}^{c}({\bf k})e^{i{\bf k}\cdot{\bf
x}}+a_{Q}^{c\,\mbox{\normalsize $\star$}}({\bf k})e^{-i{\bf k}\cdot{\bf
x}}\right]\nonumber\\
&+&\sum_{\bf k}\frac{({\sf m}_c-\bar{\sf
m}_c)^{3/2}k_l}{16k^3}\left[a_{R}^{c}({\bf k})e^{i{\bf k}\cdot{\bf
x}}+a_{R}^{c\,\mbox{\normalsize $\star$}}({\bf k})e^{-i{\bf k}\cdot{\bf
x}}\right],
\label{eq:fieldA}
\end{eqnarray}
\begin{eqnarray}
\Pi_{l}^{c}({\bf x}) &=& \sum_{\bf k} \sqrt{\frac{{\sf m}_c({\sf m}_c+\bar{\sf
m}_c)}{8\omega_c(k)}}\, \frac{ik_l}{k}\left[a^{c}({\bf k})e^{i{\bf k}\cdot{\bf
x}}-a^{c\dagger}({\bf k})e^{-i{\bf k}\cdot{\bf x}}\right]\nonumber\\
&+&\sum_{\bf k}\sqrt{\frac{\omega_c(k)({\sf m}_c+\bar{\sf m}_c)}{8{\sf m}_c}}\,
\frac{\epsilon_{ln}k_n}{k}\left[a^{c}({\bf k})e^{i{\bf k}\cdot{\bf
x}}+a^{c\dagger}({\bf k})e^{-i{\bf k}\cdot{\bf x}}\right]\nonumber\\
&+&\sum_{\bf k}\sqrt{\frac{\bar{\sf m}_c({\sf m}_c+\bar{\sf
m}_c)}{8\bar{\omega}_c(k)}}\, \frac{k_l}{k}\left[b^{c}({\bf k})e^{i{\bf
k}\cdot{\bf x}}+b^{c\dagger}({\bf k})e^{-i{\bf k}\cdot{\bf x}}\right]
\nonumber\\
&+&\sum_{\bf k}\sqrt{\frac{\bar{\omega}_c(k)({\sf m}_c+\bar{\sf
m}_c)}{8\bar{\sf m}_c}}\, \frac{i\epsilon_{ln}k_n}{k}\left[b^{c}({\bf
k})e^{i{\bf k}\cdot{\bf x}}-b^{c\dagger}({\bf k})e^{-i{\bf k}\cdot{\bf
x}}\right] \nonumber\\
&-&\sum_{\bf k}\frac{2k^3\epsilon_{ln}k_n}{\kappa_{c}(\gamma) {\sf m}_c\bar{\sf
m}_c\sqrt{{\sf m}_c-\bar{\sf m}_c}}\left[a_{Q}^{c}({\bf k})e^{i{\bf k}\cdot{\bf
x}}+a_{Q}^{c\,\mbox{\normalsize $\star$}}({\bf k})e^{-i{\bf k}\cdot{\bf
x}}\right]\nonumber\\
&+&\sum_{\bf k}\frac{({\sf m}_c-\bar{\sf m}_c)^{5/2}\epsilon_{ln}k_n}{32k^3}
\left[a_{R}^{c}({\bf k})e^{i{\bf k}\cdot{\bf x}}+a_{R}^{c\,\mbox{\normalsize
$\star$}}({\bf k})e^{-i{\bf k}\cdot{\bf x}}\right],
\label{eq:fieldPi}
\end{eqnarray}
\begin{eqnarray}
A_{0}^{c}({\bf x}) &=& -\sum_{\bf k}\frac{k}{\sqrt{2\omega_c(k){\sf m}_c({\sf
m}_c+\bar{\sf m}_c)}}\left[a^{c}({\bf k})e^{i{\bf k}\cdot{\bf
x}}+a^{c\dagger}({\bf k})e^{-i{\bf k}\cdot{\bf x}}\right]\nonumber\\
&+& \sum_{\bf k}\frac{ik}{\sqrt{2\bar{\omega}_c(k)\bar{\sf m}_c({\sf
m}_c+\bar{\sf m}_c)}}\left[b^{c}({\bf k})e^{i{\bf k}\cdot{\bf
x}}-b^{c\dagger}({\bf k})e^{-i{\bf k}\cdot{\bf x}}\right] \nonumber\\
&-&\sum_{\bf k}\frac{4k^3}{{\sf m}_c\bar{\sf m}_c({\sf m}_c-\bar{\sf
m}_c)^{3/2}} \left[a_{Q}^{c}({\bf k})e^{i{\bf k}\cdot{\bf
x}}+a_{Q}^{c\,\mbox{\normalsize $\star$}}({\bf k})e^{-i{\bf k}\cdot{\bf
x}}\right]\nonumber\\
&+&\sum_{\bf k} \frac{\kappa_{c}(\gamma)({\sf m}_c-\bar{\sf m}_c)^{3/2}}{16k^3}
\left[a_{R}^{c}({\bf k})e^{i{\bf k}\cdot{\bf x}}+a_{R}^{c\,\mbox{\normalsize
$\star$}}({\bf k})e^{-i{\bf k}\cdot{\bf x}}\right],
\label{eq:fieldAzero}
\end{eqnarray}
and
\begin{eqnarray}
G^{c}({\bf x}) & = & \sum_{\bf k}\frac{8ik^{3}}{\kappa_{c}(\gamma)({\sf
m}_c-\bar{\sf m}_c)^{3/2}}\left[a_{Q}^{c}({\bf k})e^{i{\bf k}\cdot{\bf
x}}-a_{Q}^{c\,\mbox{\normalsize $\star$}}({\bf k})e^{-i{\bf k}\cdot{\bf
x}}\right].
\label{eq:fieldG}
\end{eqnarray}
The unphysical scalar fields for $c=4,\ldots,8$ are
\begin{eqnarray}
\xi^{c}({\bf x}) & = & -\sum_{\bf k}\frac{4ik^{3}}{\kappa_{c}(\gamma)({\sf
m}_c\bar{\sf m}_c)^{1/2}({\sf m}_c-\bar{\sf m}_c)^{3/2}}\left[a_{Q}^{c}({\bf
k})e^{i{\bf
k}\cdot{\bf x}}-a_{Q}^{c\,\mbox{\normalsize $\star$}}({\bf k})e^{-i{\bf
k}\cdot{\bf x}}\right] \nonumber \\
&-& \sum_{\bf k}\frac{i({\sf m}_c\bar{\sf m}_c)^{1/2}
({\sf m}_c-\bar{\sf m}_c)^{3/2}}{16k^{3}}\left[a_{R}^{c}({\bf k})e^{i{\bf
k}\cdot{\bf x}}-a_{R}^{c\,\mbox{\normalsize $\star$}}({\bf k})e^{-i{\bf
k}\cdot{\bf x}}\right]
\label{eq:fieldXi}
\end{eqnarray}
and their canonically conjugate momenta
\begin{eqnarray}
\Pi_{\xi}^{c}({\bf x}) & = &-\sum_{\bf k}k\sqrt{\frac{\bar{\sf
m}_c}{2\omega_c(k)
({\sf m}_c+\bar{\sf m}_c)}}\left[a^{c}({\bf k})e^{i{\bf
k}\cdot{\bf x}}+a^{c\dagger}({\bf k})e^{-i{\bf k}\cdot{\bf x}}\right]
\nonumber\\
&+& \sum_{\bf k}ik\sqrt{\frac{{\sf m}_c}{2\bar{\omega}_c(k)
({\sf m}_c+\bar{\sf m}_c)}}\left[b^{c}({\bf k})e^{i{\bf k}\cdot{\bf
x}}-b^{c\dagger}({\bf k})e^{-i{\bf k}\cdot{\bf x}}\right] \nonumber \\
  &-& \sum_{\bf k}\frac{8k^{3}}{({\sf m}_c\bar{\sf m}_c)^{1/2}({\sf
m}_c-\bar{\sf m}_c)^{3/2}}\left[a_{Q}^{c}({\bf k})e^{i{\bf k}\cdot{\bf
x}}+a_{Q}^{c\,\mbox{\normalsize $\star$}}({\bf k})e^{-i{\bf k}\cdot{\bf
x}}\right],
\label{eq:fieldPiXi}
\end{eqnarray}
where $\omega_c(k)=\sqrt{{\sf m}_{c}^{2}+k^{2}}$,
$\bar{\omega}_c(k)=\sqrt{\bar{\sf m}_{c}^{2}+k^{2}}$, and
\begin{equation}
\kappa_{c}(\gamma)=\sqrt{k^{2}+(1-\gamma){\sf m}_c\bar{\sf m}_c}\,.
\end{equation}
${\sf m}_c$ and $\bar{\sf m}_c$ are the masses of $a^{c}({\bf k})$ and
$b^{c}({\bf k})$ modes, respectively. They are combinations of the Chern-Simons
topological mass $m$ and of the Higgs-Kibble mass ${\sf m}_c$; their values are
\begin{equation}
{\sf m}_c=\frac{\sqrt{4\, M_{c}^{2}+m^{2}}+ m}{2}
\label{eq:Massformula1}
\end{equation}
and
\begin{equation}
\bar{\sf m}_c=\frac{\sqrt{4\, M_{c}^{2}+m^{2}}- m}{2}.
\label{eq:Massformula2}
\end{equation}
The masses $M_{c}$ are given by Eq.~(\ref{eq:mass}). The Higgs field $\psi$ and
its canonical momentum $\Pi_{\psi}$ are represented as
\begin{equation}
\psi({\bf x})  =  \sum_{\bf k}\frac{1}{\sqrt{2\Omega(k)}}\left[\alpha({\bf
k})e^{i{\bf k}\cdot{\bf x}} +\alpha^{\dagger}({\bf k})e^{-i{\bf k}\cdot{\bf
x}}\right]
\label{eq:fieldHiggs}
\end{equation}
and
\begin{equation}
\Pi_{\psi}({\bf x})  =  -\sum_{\bf k}i\sqrt{\frac{\Omega(k)}{2}}
\left[\alpha({\bf k})e^{i{\bf k}\cdot{\bf x}}-\alpha^{\dagger}({\bf
k})e^{-i{\bf k}\cdot{\bf x}}\right],
\label{eq:fieldPiHiggs}
\end{equation}
where $\Omega(k)$ is given by
\begin{equation}
\Omega(k) = \sqrt{2{\mu}^2+k^2}.
\end{equation}
The Faddeev-Popov ghost fields are represented as \cite{KHBrick}
\begin{equation}
\sigma_{\text{f}}^{c}({\bf x})  =  \sum_{\bf
k}\frac{1}{\sqrt{2k}}\left[g_{\text{f}}^{\,c}({\bf k})e^{i{\bf
k}\cdot{\bf x}}+g_{\text{f}}^{\,c\,\mbox{\normalsize $\star$}}({\bf
k})e^{-i{\bf k}\cdot{\bf x}}\right],
\label{eq:fieldsigmaf}
\end{equation}
\begin{equation}
\sigma_{\text{p}}^{c}({\bf x})  = -\sum_{\bf
k}\frac{i}{\sqrt{2k}}\left[g_{\text{p}}^{\,c}({\bf k})e^{i{\bf
k}\cdot{\bf x}}-g_{\text{p}}^{\,c\,\mbox{\normalsize $\star$}}({\bf
k})e^{-i{\bf k}\cdot{\bf x}}\right],
\label{eq:fieldsigmap}
\end{equation}
\begin{equation}
\Pi_{\text{f}}^{c}({\bf x})  =  \sum_{\bf
k}i\sqrt{\frac{k}{2}}\left[g_{\text{p}}^{\,c}({\bf k})e^{i{\bf
k}\cdot{\bf x}}+g_{\text{p}}^{\,c\,\mbox{\normalsize $\star$}}({\bf
k})e^{-i{\bf k}\cdot{\bf x}}\right],
\label{eq:fieldPif}
\end{equation}
and
\begin{equation}
\Pi_{\text{p}}^{c}({\bf x})  =  \sum_{\bf
k}\sqrt{\frac{k}{2}}\left[g_{\text{f}}^{\,c}({\bf k})e^{i{\bf
k}\cdot{\bf x}}-g_{\text{f}}^{\,c\,\mbox{\normalsize $\star$}}({\bf
k})e^{-i{\bf k}\cdot{\bf x}}\right],
\label{eq:fieldPip}
\end{equation}
where $g_{\text{f}}^{\,c}({\bf k})$, $g_{\text{p}}^{\,c}({\bf k})$,
$g_{\text{f}}^{\,c\,\mbox{\normalsize $\star$}}({\bf k})$, and
$g_{\text{p}}^{\,c\,\mbox{\normalsize $\star$}}({\bf k})$ obey the
anticommutation rules
\begin{equation}
\{g_{\text{f}}^{\,a}({\bf k}),g_{\text{p}}^{\,b\,\mbox{\normalsize
$\star$}}({\bf q})\} = \{g_{\text{p}}^{\,a}({\bf
k}),g_{\text{f}}^{\,b\,\mbox{\normalsize $\star$}}({\bf q})\} =
\delta^{ab}\,\delta_{\bf kq}
\end{equation}
and
\begin{equation}
\{g_{\text{f}}^{\,a}({\bf k}),g_{\text{f}}^{\,b\,\mbox{\normalsize
$\star$}}({\bf q})\} = \{g_{\text{p}}^{\,a}({\bf
k}),g_{\text{p}}^{\,b\,\mbox{\normalsize $\star$}}({\bf q})\} = 0.
\end{equation}

When Eqs.~(\ref{eq:fieldA})--(\ref{eq:fieldPip}) are substituted into the
Hamiltonian $H_0$ given in Eq.~(\ref{eq:Ham0}), we obtain the expression
\begin{equation}
H_0=\sum_{c=1}^{8}H_c + H_\psi + H_{\text{fp}},
\end{equation}
where $H_c$ is given by
\begin{eqnarray}
H_{c} &=& \sum_{\bf k}\omega(k)a^{c\dagger}({\bf k})a^{c}({\bf k})+\sum_{\bf
k}k\left[a_{R}^{c\,\mbox{\normalsize $\star$}}({\bf k})a_{Q}^{c}({\bf
k})+a_{Q}^{c\,\mbox{\normalsize $\star$}}({\bf k})a_{R}^{c}({\bf
k})\right]\nonumber\\
&-&(1-\gamma)\sum_{\bf k}\frac{64k^4}{m^3}\,a_{Q}^{c\,\mbox{\normalsize
$\star$}}({\bf k})a_{Q}^{c}({\bf k})
\label{eq:HamOpunbroken}
\end{eqnarray}
for the $c=1,2,3$ sector of unbroken $SU(2)$ gluon triplet, and
\begin{eqnarray}
H_{c} & = & \sum_{\bf k}\left[\omega_c(k)a^{c\dagger}({\bf k})a^{c}({\bf
k})+\bar{\omega}_c(k)b^{c\dagger}({\bf k})b^{c}({\bf k})\right] \nonumber  \\
 & + & \sum_{\bf k}\kappa_{c}(\gamma)\left[a_{R}^{c\,\mbox{\normalsize
$\star$}}({\bf k})a_{Q}^{c}({\bf k})+a_{Q}^{c\,\mbox{\normalsize $\star$}}({\bf
k})a_{R}^{c}({\bf k})\right],
\label{eq:HamOpbroken}
\end{eqnarray}
for $c=4,\ldots,8$.  For the doublet $(c=4,5,6,7)$ and singlet $(c=8)$ sectors,
${\sf m}_{c}$ and $\bar{\sf m}_{c}$ are given by Eqs.~(\ref{eq:Massformula1})
and (\ref{eq:Massformula2}) respectively; for $c=1,2,3$ there is only a single
gluon mode and the  mass $m$ is the topological mass.  The Higgs Hamiltonian
$H_\psi$ is given by
\begin{equation}
H_{\psi}  =  \sum_{\bf k}\Omega(k){\alpha}^{\dagger}({\bf k}){\alpha}({\bf k});
\label{eq:HamHiggs}
\end{equation}
and the Faddeev-Popov ghost part of the Hamiltonian $H_{\text{fp}}$, by
\begin{equation}
H_{\text{fp}}  =  \sum_{\bf k}k\left[g_{\text{f}}^{\,c\,\mbox{\normalsize
$\star$}}({\bf k})g_{\text{p}}^{\,c}({\bf k}) +
g_{\text{p}}^{\,c\,\mbox{\normalsize $\star$}}({\bf k})g_{\text{f}}^{\,c}({\bf
k})\right].
\label{eq:HamFadPop}
\end{equation}

Inspection confirms that $H_0$ is diagonal in the particle number for the
observable, propagating particle modes (the massive gluons and the Higgs
excitations) of this model and that to this extent the representations of the
gauge fields have turned out to be ``suitable.''  Explicit construction of a
Fock space for this model will demonstrate that the ghost components of the
Hamiltonian also satisfy the suitability requirement.  We can construct a Fock
space $\{|h\rangle\}$ for this model, on the foundation of the
perturbative vacuum, $|0\rangle$, which is annihilated by all the annihilation
operators: $a^c({\bf k})$, $b^c({\bf k})$, $a_Q^c({\bf k})$ and $a_R^c({\bf
k})$, as well as ${\alpha}({\bf k})$ and the Faddeev-Popov ghosts,
$g_{\text{f}}^{\,c}({\bf k})$ and  $g_{\text{p}}^{\,c}({\bf k})$. In this
construction, we make use of techniques developed in earlier work
\cite{HLL1,HLL3,el3,khqedtemp}. This perturbative Fock space includes all
multiparticle states, $|N\rangle$, consisting of observable, propagating
particles (Higgs particles and massive gluons) that are created when
${\alpha}^{\dagger}({\bf k})$,  $a^{c\dagger}({\bf k})$ and $b^{c\dagger}({\bf
k})$ respectively act on $|0\rangle$. All such states $|N\rangle$ are
eigenstates of $H_0$. States, such as $a_Q^{c\,\mbox{\normalsize $\star$}}({\bf
k})|N\rangle$ or $a_Q^{c\,\mbox{\normalsize $\star$}}({\bf
k})a_Q^{d\,\mbox{\normalsize $\star$}}({\bf q})|N\rangle$, in which a single
variety of ghost creation operator acts on one of these multiparticle states
$|N\rangle$ have zero norm; they have no probability of being observed, and
have vanishing expectation values of energy, momentum, as well as all other
observables. We will designate as $\{|n\rangle\}$ that subspace of
$\{|h\rangle\}$ which consists of all states $|N\rangle$ and of all
states in which a chain of  $a_Q^{c\,\mbox{\normalsize $\star$}}({\bf k})$
operators\,---\,but {\em no} $a_R^{c\,\mbox{\normalsize $\star$}}({\bf k})$
operators\,---\,act on $|N\rangle$. States in which both varieties of ghosts
appear simultaneously, such as $a_Q^{c\,\mbox{\normalsize $\star$}}({\bf
k})a_R^{d\,\mbox{\normalsize $\star$}}({\bf q})|N\rangle$, are  in the Fock
space $\{|h\rangle\}$, but not in $\{|n\rangle\}$; because these states have a
nonvanishing norm and contain ghosts, they are not probabilistically
interpretable. Their appearance in the course of time
evolution signals a defect in the theory. Since the states $|N\rangle$
constitute the set of states in $\{|n\rangle\}$ from which all zero norm states
(the ones with ghost constituents) have been excised, we will sometimes speak
of the set of $|N\rangle$ as a quotient space of observable propagating states.
The time-evolution operator $\exp\left(-iH_0t\right)$ has the important
property that, if it acts on a state vector $|n_i\rangle$ in $\{|n\rangle\}$,
it can only propagate it within $\{|n\rangle\}$. We observe that the only parts
of $H_0$ that could possibly cause a state vector to leave the subspace
$\{|n\rangle\}$, are those that contain either $a_R^{c\,\mbox{\normalsize
$\star$}}({\bf k})$ or $a_R^c({\bf k})$ operators. The only part of $H_0$ that
has that feature contains the combination of operators  ${\Gamma}^c =
a_R^{c\,\mbox{\normalsize $\star$}}({\bf k})a_Q^c({\bf k}) +
a_Q^{c\,\mbox{\normalsize $\star$}}({\bf k})a_R^c({\bf k})$. When $a_R^c({\bf
k})$ acts on a state vector $|n_i\rangle$, it either annihilates the vacuum or
annihilates one of the $a_Q^{c\,\mbox{\normalsize $\star$}}({\bf k})$ operators
in $\{|n\rangle\}$. In the latter case, ${\Gamma}^c$ replaces the annihilated
$a_Q^{c\,\mbox{\normalsize $\star$}}({\bf k})$ operator with an identical one.
When $a_Q^c({\bf k})$ acts on a state vector $|n_i\rangle$, it always
annihilates it. It is therefore impossible for ${\Gamma}^c$ to transform a
state vector in $\{|n\rangle\}$ to  one external to $\{|n\rangle\}$ in which an
$a_R^{c\,\mbox{\normalsize $\star$}}({\bf k})$ operator acts on $|n_i\rangle$.
The only effect of ${\Gamma}^c$ is to translate $|n_i\rangle$ states within
$\{|n\rangle\}$.  These features of the Hamiltonian $H_0$ confirm that
Eqs.~(\ref{eq:fieldA})--(\ref{eq:fieldPiXi}) are suitable representations of
the gauge fields.  $H_0$ counts the number of massive gluons of momentum ${\bf
k}$  belonging to  the unbroken $SU(2)$ sector of the original $SU(3)$ system,
and assigns an energy  $\omega(k)$ to each of them.  It similarly counts the
two  varieties of massive gluons in the doublet and singlet sectors, and
assigns the  energy ${\omega}_c$ and $\bar{\omega}_c(k)$ to the $a^c({\bf k})$
and $b^c({\bf k})$ varieties respectively.  And lastly, $H_0$ counts the number
of Higgs particles of mass $\sqrt{2}\mu$  and assigns the energy
$\displaystyle\sqrt{2{\mu}^{2}+k^2}$ to each.  Beyond that, the form of $H_0$
guarantees that any state vector initially in $\{|n\rangle\}$  is propagated by
$\exp\left(-iH_0t\right)$ entirely within $\{|n\rangle\}$.

We next turn to the implementation of Gauss's law and the gauge condition.  We
have previously noted that Gauss's law, ${\cal G}^a({\bf x})=0$, is not a
consequence of the Euler-Lagrange equations, and that further analysis is
required to demonstrate that it is properly implemented. We further observe
that, when Eqs.~(\ref{eq:fieldA})--(\ref{eq:fieldPip}) are substituted into
Eq.~(\ref{eq:Gscript}), ${\cal G}^a$ turns out to be a linear combination of
only those ghost excitations that can live in the subspace
$\{|n\rangle\}$\,---\,$a_Q^a({\bf k})$ and $a_Q^{a\mbox{\normalsize
$\star$}}({\bf k})$. All other excitation operators\,---\,$a_R^a({\bf k})$ and
$a_R^{a\mbox{\normalsize $\star$}}({\bf k})$, and the annihilation and creation
operators for both varieties of propagating particles which appear in the gauge
fields, their canonical momenta, and in $\Pi_\xi^a$\,---\,cancel in ${\cal
G}^a$. The explicit expression for ${\cal G}^a$ obtained from this substitution
is
\begin{equation}
{\cal G}^a({\bf x}) = \sum_{\bf k}\frac{8k^3}{(\sf{m}-\bar{\sf
m})^{3/2}}\left[a_{Q}^{a}({\bf k})e^{i{\bf k}\cdot{\bf
x}}+a_{Q}^{a\,\mbox{\normalsize $\star$}}({\bf k})e^{-i{\bf k}\cdot{\bf
x}}\right].
\end{equation}
The implementation of Gauss's law is an immediate consequence of this
expression for ${\cal G}^a$. A state vector that describes an observable state
is one of the $|N\rangle$ states in the quotient space discussed earlier. The
time evolution generated by $\exp\left(-iH_0t\right)$ has previously been shown
to keep any state vector that initially was an $|N\rangle$ state contained in
the subspace $\{|n\rangle\}$. And the Gauss's law operator ${\cal G}^a$, as
well as any other operator that is a linear combination of $a_Q^a({\bf k})$ and
$a_Q^{a\mbox{\normalsize $\star$}}({\bf k})$ operators, must vanish in
$\{|n\rangle\}$. These facts provide for the permanent validity of Gauss's law
as long as the state vector representing the system is initially one of the
$|N\rangle$ state\,---\,or at least a state in $\{|n\rangle\}$\,---\,and
provided that $\exp\left(-iH_0t\right)$ is the time-evolution operator for the
system. Similarly, $G^c$ is also represented as a superposition of $a_Q^c({\bf
k})$ and $a_Q^{c\mbox{\normalsize $\star$}}({\bf k})$ ghost excitation
operators only, so that $\langle n_b|G^c|n_a\rangle = 0$ for the same reason
that $\langle n_b|{\cal G}^c|n_a\rangle = 0$.
Equation~(\ref{eq:gaugecondition}) therefore shows that in the subspace
$\{|n\rangle\}$, the t'Hooft gauge condition, $\partial_\mu
A^{a\mu}-(1-\gamma)\alpha^a = 0$, holds. We thus have shown not only that the
time-displacement operator $\exp\left(-iH_0t\right)$ keeps state vectors
permanently within the subspace $\{|n\rangle\}$, but that it is also precisely
in this subspace that Gauss's law and the gauge condition are permanently
implemented.

It is apparent that the explicit representations of the fields we have given in
Eqs.~(\ref{eq:fieldA})--(\ref{eq:fieldPiXi}) are instrumental in obtaining the
results we have demonstrated above. But the confirmation of the particle mode
content of these fields that the self-consistency of this formulation provides
is not weakened by its dependence on an explicit representation of the fields.
A representation in terms of creation and annihilation operators, and the
choice of a Hilbert space in which to embed the formalism\,---\,in this case
the Fock spaces $\{|n\rangle\}$ and $\{|h\rangle\}$\,---\,are inevitably
important parts of the axiomatic structure of the theory.  And it is a
significant fact that a representation of the operator-valued fields and a Fock
space have been found that permit a consistent interpretation  of $H_0$ as a
kinetic energy operator for a system of noninteracting particles  in a new
vacuum state, even though part of the interaction  described by ${\cal L}$ is
included in $H_0$. Moreover, a Fock space has been constructed within which
$H_0$ time displaces state vectors so that  unitarity, Gauss's law, and the
gauge condition are all permanently guaranteed. It should be noted that when
{\it all\/} interactions are  included in a {\it complete\/} Hamiltonian $H$,
these conditions no longer apply.  Under the influence of the time-evolution
operator $\exp(-iHt)$, state vectors  ``leak out'' of $\{|n\rangle\}$, and
probabilistically uninterpretable state vectors that contain combination of
ghosts, for example $a_Q^{c\,\mbox{\normalsize $\star$}}({\bf
k})a_R^{d\,\mbox{\normalsize $\star$}}({\bf q})|N\rangle$, develop.
Combinations of Faddeev-Popov ghosts are then necessary to compensate for  such
combinations of $Q$ and $R$ ghosts \cite{KHBrick}, and loops of Faddeev-Popov
ghost play an important role in maintaining the unitarity of the theory. One
reason for the interpretability of this model is that  the ``interaction-free''
 limit we have described\,---\,the limit as $e\rightarrow0$ and $h\rightarrow0$
while $e^2/h$ remains constant\,---\,leads to an  essentially Abelian theory.
The fact that
$[{\cal G}^{a}({\bf x}),{\cal G}^{b}({\bf y})]=0$ confirms that observation.
In a non-Abelian theory this commutator would not vanish, but would regenerate
the Gauss's law operator ${\cal G}^{c}({\bf x})$ in a pattern determined by the
structure constants of the corresponding Lie group. Because of the Abelian
nature of this limiting form of the theory, the Faddeev-Popov ghost are not
required in this stage of the work, and have not been included in the Fock
space $\{|h\rangle\}$ or $\{|n\rangle\}$.

\section{The perturbative theory}
The propagator for the gauge field is given by
\begin{equation}
D_{\mu\nu}(x_{1},x_{2})=\langle 0|{\sf
T}[A_{\mu}(x_{1}),A_{\nu}(x_{2})]|0\rangle,
\label{eq:propagatorgauge}
\end{equation}
where {\sf T} designates time-ordering, $A_{\mu}(x)$ is the interaction-picture
field
\begin{equation}
A_{\mu}(x) = e^{iH_{0}t}A_{\mu}({\bf x})e^{-iH_{0}t},
\label{eq:intpicture}
\end{equation}
$A_{\mu}({\bf x})$ is the Schr\"odinger picture field, and $|0\rangle$ is the
vacuum state of the $\{|n\rangle\}$ space.  Similarly, the propagator for an
unphysical scalar ${\xi}(x)$ is
\begin{equation}
\Delta_\xi(x_{1},x_{2})=\langle 0|{\sf T}[\xi(x_{1}),\xi(x_{2})]|0 \rangle,
\label{eq:propagatorscalar}
\end{equation}
and, for the Higgs field,
\begin{equation}
\Delta_\psi(x_{1},x_{2})=\langle 0|{\sf T}[\psi(x_{1}),\psi(x_{2})]|0 \rangle.
\label{eq:propagatorHiggs}
\end{equation}
There are other propagators in this theory, but they vanish for ${\gamma}=1$
(Landau gauge)  which we use in our work, and therefore are not of primary
interest to us.  We find that the relevant interaction picture fields for
$c=1,2,3$ are
\begin{eqnarray}
A_l^{c}(x) &=& \sum_{\bf k}\frac{8ik\epsilon_{ln}k_n}{m^{5/2}}\left[a_Q^c({\bf
k})e^{i{\bf k\cdot x}-ikt}-a_Q^{c\,\mbox{\normalsize $\star$}}({\bf
k})e^{-i{\bf k\cdot x}+ikt}\right]\nonumber\\
&+& (1-\gamma)\sum_{\bf k}\frac{2k_l}{m^{3/2}}\left[a_Q^c({\bf k})e^{i{\bf
k\cdot x}-ikt}+a_Q^{c\,\mbox{\normalsize $\star$}}({\bf k})e^{-i{\bf k\cdot
x}+ikt}\right]\nonumber\\
&-&\sum_{\bf k}\frac{4k^2k_l}{m^{7/2}}\left[a_Q^c({\bf k})e^{i{\bf k\cdot
x}-ikt}+a_Q^{c\,\mbox{\normalsize $\star$}}({\bf k})e^{-i{\bf k\cdot
x}+ikt}\right]\nonumber\\
&+&\sum_{\bf k}\frac{m^{3/2}k_l}{16k^3}\left[a_R^c({\bf k})e^{i{\bf k\cdot
x}-ikt}+a_R^{c\,\mbox{\normalsize $\star$}}({\bf k})e^{-i{\bf k\cdot
x}+ikt}\right]\nonumber\\
&-&\sum_{\bf k}\frac{\sqrt{\omega(k)}k_l}{\sqrt{2}mk}\left[a^c({\bf k})e^{i{\bf
k\cdot x}-i\omega(k)t}+a^{c\dagger}({\bf k})e^{-i{\bf k\cdot
x}+i\omega(k)t}\right]\nonumber\\
&+&\sum_{\bf k}\frac{i\epsilon_{ln}k_n}{k\sqrt{2\omega(k)}}\left[a^c({\bf
k})e^{i{\bf k\cdot x}-i\omega(k)t}-a^{c\dagger}({\bf k})e^{-i{\bf k\cdot
x}+i\omega(k)t}\right]
\label{eq:Ailt}
\end{eqnarray}
and
\begin{eqnarray}
A_0^{c}({\bf x}) &=& -\sum_{\bf k}\frac{4k^3}{m^{7/2}}\left[a_Q^c({\bf
k})e^{i{\bf k\cdot x}-ikt}+a_Q^{c\,\mbox{\normalsize $\star$}}({\bf
k})e^{-i{\bf k\cdot x}+ikt}\right]\nonumber\\
&-&(1-\gamma)\sum_{\bf k}\frac{2k}{m^{3/2}}\left[a_Q^c({\bf k})e^{i{\bf k\cdot
x}-ikt}+a_Q^{c\,\mbox{\normalsize $\star$}}({\bf k})e^{-i{\bf k\cdot
x}+ikt}\right]\nonumber\\
&+&\sum_{\bf k}\frac{m^{3/2}}{16k^2}\left[a_R^c({\bf k})e^{i{\bf k\cdot
x}-ikt}+a_R^{c\,\mbox{\normalsize $\star$}}({\bf k})e^{-i{\bf k\cdot
x}+ikt}\right]\nonumber\\
&-&\sum_{\bf k}\frac{k}{m\sqrt{2\omega(k)}}\left[a^c({\bf k})e^{i{\bf k\cdot
x}-i\omega(k)t}+a^{c\dagger}({\bf k})e^{-i{\bf k\cdot x}+i\omega(k)t}\right];
\end{eqnarray}
for $c=4,\ldots,8$, they are
\begin{eqnarray}
A_{l}^{c}(x) &=& -\sum_{\bf k} \sqrt{\frac{\omega_c(k)}{2{\sf m}_c({\sf m}_c+
\bar{\sf m}_c)}}\,\frac{k_l}{k}\left[a^{c}({\bf k})
e^{i{\bf k}\cdot{\bf x}-i\omega_c(k)t}+a^{c\dagger}({\bf k})e^{-i{\bf
k}\cdot{\bf x}+i\omega_c(k)t}\right]\nonumber\\
&+& \sum_{\bf k}\sqrt{\frac{{\sf m}_c}{2\omega_c(k)({\sf m}_c+\bar{\sf
m}_c)}}\,\frac{i\epsilon_{ln}k_n}{k}\left[a^{c}({\bf k})e^{i{\bf k}\cdot{\bf
x}-i\omega_c(k)t}-a^{c\dagger}({\bf k})e^{-i{\bf k}\cdot{\bf
x}+i\omega_c(k)t}\right]\nonumber\\
&+&\sum_{\bf k}\sqrt{\frac{\bar{\omega}_c(k)}{2\bar{\sf m}_c({\sf m}_c+\bar{\sf
m}_c)}}\, \frac{ik_l}{k}\left[b^{c}({\bf k})e^{i{\bf k}\cdot{\bf
x}-i\bar{\omega}_c(k)t}-b^{c\dagger}({\bf k})e^{-i{\bf k}\cdot{\bf
x}+i\bar{\omega}_c(k)t}\right] \nonumber\\
&-&\sum_{\bf k}\sqrt{\frac{\bar{\sf m}_c}{2\bar{\omega}_c(k)({\sf m}_c+\bar{\sf
m}_c)}}\,\frac{\epsilon_{ln}k_n}{k}\left[b^{c}({\bf k})e^{i{\bf k}\cdot{\bf
x}-i\bar{\omega}_c(k)t}+b^{c\dagger}({\bf k})e^{-i{\bf k}\cdot{\bf
x}+i\bar{\omega}_c(k)t}\right] \nonumber\\
&-& \sum_{\bf k} \frac{4k^3k_l}{\kappa_{c}(\gamma){\sf m}_c\bar{\sf m}_c({\sf
m}_c-\bar{\sf m}_c)^{3/2}} \left[a_{Q}^{c}({\bf k})e^{i{\bf k}\cdot{\bf
x}-i\kappa_{c}t}+a_{Q}^{c\,\mbox{\normalsize $\star$}}({\bf k})e^{-i{\bf
k}\cdot{\bf x}+i\kappa_{c}t}\right]\nonumber\\
&+&\sum_{\bf k}\frac{({\sf m}_c-\bar{\sf
m}_c)^{3/2}k_l}{16k^3}\left[a_{R}^{c}({\bf k})e^{i{\bf k}\cdot{\bf
x}-i\kappa_{c}t}+a_{R}^{c\,\mbox{\normalsize $\star$}}({\bf k})e^{-i{\bf
k}\cdot{\bf x}+i\kappa_{c}t}\right],
\end{eqnarray}
\begin{eqnarray}
A_{0}^{c}(x) &=& -\sum_{\bf k}\frac{k}{\sqrt{2\omega_c(k){\sf m}_c({\sf
m}_c+\bar{\sf m}_c)}}\left[a^{c}({\bf k})e^{i{\bf k}\cdot{\bf
x}-i\omega_c(k)t}+a^{c\dagger}({\bf k})e^{-i{\bf k}\cdot{\bf
x}+i\omega_c(k)t}\right]\nonumber\\
&+& \sum_{\bf k}\frac{ik}{\sqrt{2\bar{\omega}_c(k)\bar{\sf m}_c({\sf
m}_c+\bar{\sf m}_c)}}\left[b^{c}({\bf k})e^{i{\bf k}\cdot{\bf
x}-i\bar{\omega}_c(k)t}-b^{c\dagger}({\bf k})e^{-i{\bf k}\cdot{\bf
x}+i\bar{\omega}_c(k)t}\right] \nonumber\\
&-&\sum_{\bf k}\frac{4k^3}{{\sf m}_c\bar{\sf m}_c({\sf m}_c-\bar{\sf
m}_c)^{3/2}} \left[a_{Q}^{c}({\bf k})e^{i{\bf k}\cdot{\bf
x}-i\kappa_{c}t}+a_{Q}^{c\,\mbox{\normalsize $\star$}}({\bf k})e^{-i{\bf
k}\cdot{\bf x}+i\kappa_{c}t}\right]\nonumber\\
&+&\sum_{\bf k} \frac{\kappa_{c}(\gamma)({\sf m}_c-\bar{\sf m}_c)^{3/2}}{16k^3}
\left[a_{R}^{c}({\bf k})e^{i{\bf k}\cdot{\bf
x}-i\kappa_{c}t}+a_{R}^{c\,\mbox{\normalsize $\star$}}({\bf k})e^{-i{\bf
k}\cdot{\bf x}-i\kappa_{c}t}\right],
\end{eqnarray}
\begin{eqnarray}
\xi^{c}(x) & = & -\sum_{\bf k}\frac{4ik^{3}}{\kappa_{c}(\gamma)({\sf
m}_c\bar{\sf m}_c)^{1/2}({\sf m}_c-\bar{\sf m}_c)^{3/2}}\left[a_{Q}^{c}({\bf
k})e^{i{\bf k}\cdot{\bf x}-i\kappa_{c}t} - a_{Q}^{c\,\mbox{\normalsize
$\star$}}({\bf k})e^{-i{\bf k}\cdot{\bf x}+i\kappa_{c}t}\right] \nonumber \\
&-& \sum_{\bf k}\frac{i({\sf m}_c\bar{\sf m}_c)^{1/2}
({\sf m}_c-\bar{\sf m}_c)^{3/2}}{16k^{3}}\left[a_{R}^{c}({\bf k})e^{i{\bf
k}\cdot{\bf x}-i\kappa_{c}t}-a_{R}^{c\,\mbox{\normalsize $\star$}}({\bf
k})e^{-i{\bf k}\cdot{\bf x}+i\kappa_{c}t}\right],
\end{eqnarray}
and
\begin{equation}
\psi(x) = \sum_{\bf k}\frac{1}{\sqrt{2\Omega(k)}}\left[\alpha({\bf k})e^{i{\bf
k}\cdot{\bf x}-i\Omega(k)t} +\alpha^{\dagger}({\bf k})e^{-i{\bf k}\cdot{\bf
x}+i\Omega(k)t}\right].
\end{equation}
The Faddeev-Popov ghost fields are
\begin{equation}
\sigma_{\text{f}}^{c}(x)  =  \sum_{\bf
k}\frac{1}{\sqrt{2k}}\left[g_{\text{f}}^{\,c}({\bf k})e^{i{\bf k}\cdot{\bf
x}-ikt}-g_{\text{f}}^{\,c\,\mbox{\normalsize $\star$}}({\bf k})e^{-i{\bf
k}\cdot{\bf x}+ikt}\right]
\end{equation}
and
\begin{equation}
\sigma_{\text{p}}^{c}(x)  = \sum_{\bf
k}\frac{i}{\sqrt{2k}}\left[g_{\text{p}}^{\,c}({\bf k})e^{i{\bf k}\cdot{\bf
x}-ikt}-g_{\text{p}}^{\,c\,\mbox{\normalsize $\star$}}({\bf k})e^{-i{\bf
k}\cdot{\bf x}+ikt}\right].
\end{equation}
The propagators for the gauge fields can be expressed as
\begin{equation}
D_{\mu\nu}^{ab}(x_{1},x_{2})  =  -i\delta^{ab}\int\frac{d^{3}k}{(2\pi)^{3}}\
D_{\mu\nu}^{(a)}(k)e^{-ik^{\alpha}(x_{1}-x_{2})_{\alpha}};
\label{eq:prop1}
\end{equation}
for $a=1,2,3$:
\begin{eqnarray}
D^{(a)}_{\mu\nu}(k) &=& (1-\gamma)\,\frac{k_\mu k_\nu}{(k^\alpha k_\alpha +
i\epsilon)^2} - \frac{k_\mu k_\nu}{(k^\alpha k_\alpha + i\epsilon)(k^\alpha
k_\alpha - m^2 + i\epsilon)}\nonumber\\
&+& \frac{g_{\mu\nu}}{k^\alpha k_\alpha - m^2 + i\epsilon} +
\frac{im\epsilon_{\mu\nu\lambda}k^\lambda}{(k^\alpha k_\alpha +
i\epsilon)(k^\alpha k_\alpha - m^2 + i\epsilon)};
\label{eq:DmunuMCS}
\end{eqnarray}
and for $a=4,\ldots,8$:
\begin{eqnarray}
D_{\mu\nu}^{(a)}(k) &=& \frac{(k^{\alpha}k_{\alpha}-{\sf m}_a\bar{\sf
m}_a)g_{\mu\nu}}{(k^{\alpha}k_{\alpha}-{\sf m}_a^{2} +
i\epsilon)(k^{\alpha}k_{\alpha}-\bar{\sf m}_a^{2}+i\epsilon)}+\frac{i({\sf
m}_a-\bar{\sf m}_a)\epsilon_{\mu\nu\rho}k^{\rho}}{(k^{\alpha}k_{\alpha}-{\sf
m}_a^{2} + i\epsilon)(k^{\alpha}k_{\alpha}-\bar{\sf m}_a^{2} +
i\epsilon)}\nonumber\\
&-& \displaystyle\frac{\gamma(k^{\alpha}k_{\alpha}-{\sf m}_a\bar{\sf
m}_a)k_{\mu}k_{\nu}}{[k^{\alpha}k_{\alpha}-(1-\gamma){\sf m}_a\bar{\sf m}_a +
i\epsilon](k^{\alpha}k_{\alpha}-{\sf m}_a^{2} +
i\epsilon)(k^{\alpha}k_{\alpha}-\bar{\sf m}_a^{2} + i\epsilon)}\nonumber\\
&-& \frac{(1-\gamma)({\sf m}_a-\bar{\sf
m}_a)^{2}k_{\mu}k_{\nu}}{[k^{\alpha}k_{\alpha}-(1-\gamma){\sf m}_a\bar{\sf m}_a
+ i\epsilon](k^{\alpha}k_{\alpha}-{\sf m}_a^{2} +
i\epsilon)(k^{\alpha}k_{\alpha}-\bar{\sf m}_a^{2} + i\epsilon)}.
\end{eqnarray}
These expressions agree with the gauge field propagators reported in
Ref.~\cite{PR} for $a=1,2,3$ and with Refs.~\cite{PR1,hlousek,Khare} for
$a=4,\ldots,8$. These propagators were obtained by inverting the quadratic part
of the gauge-fixed Lagrangian. The other propagators are given in terms of the
Fourier integral
\begin{equation}
\Delta(x_{1},x_{2}) = -i\int\frac{d^{3}k}{(2\pi)^{3}}\
\Delta(k^2)e^{-ik^{\alpha}(x_{1}-x_{2})_{\alpha}},
\end{equation}
where
\begin{equation}
\Delta_{\psi}(k^2) = \frac{-1}{k^{\alpha}k_{\alpha}-2{\mu}^{2} + i\epsilon},
\end{equation}
\begin{equation}
\Delta_{\xi}^{(a)}(k^2) =
\frac{-\delta^{ab}}{k^{\alpha}k_{\alpha}-(1-\gamma){\sf m}_a\bar{\sf m}_a +
i\epsilon},
\label{eq:prop2}
\end{equation}
and
\begin{equation}
\Delta_{\text{fp}}(k^2) = \frac{-1}{k^\alpha k_\alpha + i\epsilon}.
\end{equation}

In a canonical theory, the vertices are dictated by the interaction Hamiltonian
$H_{\text{int}}$.  Since, in this model, time derivatives of operator-valued
fields appear in the interaction Lagrangian as well as in ${\cal L}_{0}$,
$H_{\text{int}}$ will differ from $-\int d{\bf x}\ ({\cal L}_{1}+{\cal
L}_{2})$.  The resulting vertices will be determined by $H_{\text{int}}$, and
the propagators will consist of vacuum expectation values of the time-ordered
fields that appear in  $H_{\text{int}}$.  In expanding the $S$-matrix for
scattering from an initial state $|i\rangle$ to a final state $|f\rangle$,
\begin{equation}
S_{fi}= \left\langle f\left|{\sf T} \exp\left(-i\int dt\ e^{iH_{0}t}
H_{\text{int}} e^{-iH_{0}t}\right)\right|i\right\rangle,
\end{equation}
by using the Wick theorem \cite{Wick}, we will sometimes encounter time-ordered
products of fields and, at other times, time-ordered products of space-time
derivatives of fields.  When time derivatives of fields appear as arguments of
a time-ordering operation, we will replace the time-ordering operator ${\sf T}$
with the ``{\sf T}-star ordering'' operator ${\sf T}^{\mbox{\normalsize
$\ast$}}$ which is defined so that any derivatives acting on time-ordered
fields are to be taken only {\it after\/} time ordering has been carried out.
In transforming ${\sf T}$-ordered to ${\sf T}^{\mbox{\normalsize
$\ast$}}$-ordered fields, additional terms are generated, which contain the
${\delta}(x_{0}-y_{0})$ that is produced when time derivatives are extracted
from ${\sf T}$-ordered products of time-differentiated fields. As was pointed
out by Matthews, these extra terms in which ${\delta}(x_{0}-y_{0})$-functions
appear just cancel the difference between $H_{\text{int}}$ and $-\int d{\bf x}
\ ({\cal L}_{1}+{\cal L}_{2})$, so that the perturbative theory requires only
the propagators given in Eqs.~(\ref{eq:prop1})--(\ref{eq:prop2}) and the
vertices dictated by the interaction Lagrangian \cite{Matthews}.  Application
of the Matthews rule to a model with a spontaneously broken gauge symmetry that
produces massive gauge excitations also applies to this case \cite{KH1}.

\section{Poincar\'e structure and Lorentz transformations of massive gauge
bosons}
In this section we will construct the six canonical Poincar\'e generators in
$2+1$ dimensions: the time-evolution operator, $P_0=H_0$; the two-component
space-displacement operator $P_l$; the (scalar) rotation operator $J$; and the
two-component Lorentz boost $K_l$.  We will also use the Lorentz boost
generators to transform the single-particle massive gauge boson states, to
display their properties under Lorentz transformations as well as to obtain
further confirmation of the consistency of our canonical formulation of this
model.

The canonical Poincar\'e generators for this model are: $P_0 = \int d{\bf x}\
{\cal P}_0({\bf x})$, where ${\cal P}_0={\cal H}_0$ with ${\cal H}_0$ given by
Eq.~(\ref{eq:Ham0});
\begin{equation}
P_l = \int d{\bf x}\ {\cal P}_l({\bf x})
\label{eq:Pl}
\end{equation}
where
\begin{equation}
{\cal P}_l = -\Pi_\xi\partial_l\xi - \Pi_n\partial_lA_n + G\partial_lA_0 -
\Pi_\psi\partial_l\psi - \Pi_{\text{f}}\partial_l\sigma_{\text{f}} -
\Pi_{\text{p}}\partial_l\sigma_{\text{p}};
\end{equation}
\begin{equation}
J = \int d{\bf x}\ \epsilon_{ln}x_l{\cal P}_n({\bf x}) + \int d{\bf x}\
\kappa_{\text{rotation}}({\bf x})
\label{eq:J}
\end{equation}
and
\begin{equation}
K_l = x_0P_l - \int d{\bf x}\ x_l{\cal P}_0({\bf x}) + \int d{\bf x}\
\kappa^{\text{boost}}_l({\bf x})
\label{eq:Kl}
\end{equation}
where
\begin{equation}
\kappa_{\text{rotation}} = \epsilon_{ln}A_l\Pi_n
\end{equation}
and
\begin{equation}
\kappa^{\text{boost}}_l = -A_lG + A_0\Pi_l.
\end{equation}
The term $\kappa_{\text{rotation}}$ implements the mixing of the space
components of the fields during a rotation. It arises from the fact that, under
an infinitesimal rotation $\delta\theta$ about an axis perpendicular to the 2-D
plane, the components of $A^\mu$ transform as follows:
\begin{equation}
\delta A_l({\bf x}) = -[\epsilon_{ij}x_i\partial_jA_l({\bf x}) +
\epsilon_{ln}A_n({\bf x})]\,\delta\theta
\label{eq:deltaAl}
\end{equation}
and
\begin{equation}
\delta A_0({\bf x}) = -\epsilon_{ij}x_i\partial_jA_0({\bf x})\,\delta\theta.
\end{equation}
Under an infinitesimal boost $\delta\beta_l$ along the $l$-direction, the
components of $A^\mu$ transform as follows
\begin{equation}
\delta A_0({\bf x}) = -[x_0\partial_lA_0({\bf x}) + x_l\partial_0A_0({\bf x})-
A_l({\bf x})]\,\delta\beta_l
\label{eq:deltaA0betal}
\end{equation}
and
\begin{equation}
\delta A_i({\bf x}) = -[x_0\partial_lA_i({\bf x}) + x_l\partial_0A_i({\bf x})-
\delta_{il}A_0({\bf x})]\,\delta\beta_l.
\label{eq:deltaAibetal}
\end{equation}

Use of the canonical commutation rules leads to the following commutation rules
for the Poincar\'e generators:
\begin{equation}
[P_l,P_n]=0,
\label{eq:PlPn}
\end{equation}
\begin{equation}
[H,P_l] = [H,J]=0,
\label{eq:PlJH}
\end{equation}
\begin{equation}
[H,K_l]=iP_l,
\label{eq:HKl}
\end{equation}
\begin{equation}
[P_l,K_n] = i\delta_{ln}H,
\label{eq:PlKn}
\end{equation}
\begin{equation}
[P_l,J] = -i\epsilon_{ln}P_n,
\label{eq:PlJ}
\end{equation}
\begin{equation}
[J,K_l] = i\epsilon_{ln}K_n,
\label{eq:JKl}
\end{equation}
and
\begin{equation}
[K_l,K_n] = -i\epsilon_{ln}J.
\label{eq:KlKn}
\end{equation}
We observe that these commutation rules form a closed Lie algebra, and that
they are consistent with the transformations given in
Eqs.~(\ref{eq:deltaAl})--(\ref{eq:deltaAibetal}).

To facilitate this investigation of the Lorentz transformation of states that
are eigenstates to ${H}_0$, we shift to a description of excitation operators
that have an invariant norm under Lorentz transformations. We observe, for
example, that the norm of the one-particle state  $a^{c\dagger}({\bf
k})|0\rangle$,
\begin{equation}
\left|a^{c\dagger}({\bf k})|0\rangle\right|^2 = \sum_{{\bf q}}\langle
0|[a^c({\bf q}),a^{c\dagger}({\bf k})]|0\rangle = \int d{\bf q}\ \delta({\bf
k-q}),
\end{equation}
is not a Lorentz scalar because $d{\bf k}$ is not the Lorentz invariant measure
for the phase space. The invariant measure can be established by noting that
the invariant delta function
\begin{equation}
\delta({\bf k}-{\bf q})\delta(k_0-q_0)\delta(q_\mu q^\mu - {\sf
m}_c^2)\Theta(q_0) = \frac{\delta({\bf k}-{\bf
q})\delta(k_0-\omega_c(k))}{2\omega_c(k)},
\end{equation}
so that the states $A^{c\dagger}({\bf k})|0\rangle$, created by operators that
obey
\begin{equation}
[A^c({\bf k}),A^{d\dagger}({\bf q})] =
2\omega_c(k)(2\pi)^2\delta^{cd}\,\delta({\bf k}-{\bf q}),
\label{eq:AAdagger}
\end{equation}
have unit norms in every Lorentz frame. Similarly, the normalized operators for
the other modes of the gauge field obey
\begin{equation}
[B^{\,c}({\bf k}),B^{\,d\dagger}({\bf q})] =
2\bar{\omega}_c(k)(2\pi)^2\delta^{cd}\,\delta({\bf k}-{\bf q});
\label{eq:BBdagger}
\end{equation}
and the equivalently normalized ghost operators satisfy
\begin{equation}
[A_Q^c({\bf k}),A_R^{d\,\mbox{\normalsize $\star$}}({\bf q})] = [A_R^c({\bf
k}),A_Q^{d\,\mbox{\normalsize $\star$}}({\bf q})] =
2\kappa_c(\gamma)(2\pi)^2\delta^{cd}\,\delta({\bf k}-{\bf q}).
\label{eq:ARAQdagger}
\end{equation}
The normalized operators corresponding to the mode $\alpha({\bf k})$ of the
Higgs field and the two Faddeev-Papov ghosts $g^{\,a}_{\text{f}}({\bf k})$ and
$g^{\,a}_{\text{p}}({\bf k})$ are given by $\hat{\alpha}({\bf k})$,
$\hat{g}^{\,a}_{\text{f}}({\bf k})$ and $\hat{g}^{\,a}_{\text{p}}({\bf k})$,
respectively. These normalized operators satisfy the following commutation and
anticommutation relations:
\begin{equation}
[\hat{\alpha}({\bf k}),\hat{\alpha}^\dagger({\bf q})] =
2{\Omega}_k(2\pi)^2\,\delta({\bf k}-{\bf q})
\end{equation}
and
\begin{equation}
\{\hat{g}_{\text{f}}^{\,a}({\bf k}),\hat{g}_{\text{p}}^{\,b\,\mbox{\normalsize
$\star$}}({\bf q})\} = \{\hat{g}_{\text{p}}^{\,a}({\bf
k}),\hat{g}_{\text{f}}^{\,b\,\mbox{\normalsize $\star$}}({\bf q})\} =
2k(2\pi)^2\delta^{ab}\,\delta({\bf k}-{\bf q}).
\end{equation}

Hence, the boost operator $K_l$ is written as
\begin{eqnarray}
K_l &=& \sum_{c=1}^8\sum_{\bf k} \frac{{\sf
m}_c\epsilon_{ln}k_n}{2k^2\omega_c(k)}\,A^{c\dagger}({\bf k})A^c({\bf k}) -
\sum_{c=4}^8\sum_{\bf k} \frac{\bar{\sf
m}_c\epsilon_{ln}k_n}{2k^2\bar{\omega}_c(k)}\,B^{\,c\dagger}({\bf
k})B^{\,c}({\bf k})\nonumber\\[3pt]
&+&\sum_{c=1}^8\sum_{\bf k}\frac{i}{4}\left[\frac{\partial}{\partial
k_l}\,A^{c\dagger}({\bf k})A^c({\bf k}) - A^{c\dagger}({\bf
k})\frac{\partial}{\partial k_l}\,A^c({\bf k})\right]\nonumber\\[3pt]
&+&\sum_{c=4}^8\sum_{\bf k}\frac{i}{4}\left[\frac{\partial}{\partial
k_l}\,B^{\,c\dagger}({\bf k})B^{\,c}({\bf k}) - B^{\,c\dagger}({\bf
k})\frac{\partial}{\partial k_l}\,B^{\,c}({\bf k})\right]\nonumber\\[3pt]
&+&\sum_{c=4}^8\sum_{\bf k}\frac{i}{4}\left[\frac{\partial}{\partial
k_l}\,\hat{\alpha}^{\,c\dagger}({\bf k})\hat{\alpha}^{\,c}({\bf k}) -
\hat{\alpha}^{\,c\dagger}({\bf k})\frac{\partial}{\partial
k_l}\,\hat{\alpha}^{\,c}({\bf k})\right]\nonumber\\[3pt]
&+& \sum_{c=1}^8\sum_{\bf k} \frac{i}{2}\left[\frac{\partial}{\partial
k_l}\,A_Q^{c\,\mbox{\normalsize $\star$}}({\bf k})A_R^c({\bf k}) -
A_R^{c\,\mbox{\normalsize $\star$}}({\bf k})\frac{\partial}{\partial
k_l}\,A_Q^c({\bf k})\right]\nonumber\\[3pt]
&+& \sum_{c=1}^8\sum_{\bf k} \frac{i}{2}\left[\frac{\partial}{\partial
k_l}\,\hat{g}_{\text{f}}^{\,c\,\mbox{\normalsize $\star$}}({\bf
k})\hat{g}^{\,c}_{\text{p}}({\bf k}) -
\hat{g}_{\text{p}}^{\,c\,\mbox{\normalsize $\star$}}({\bf
k})\frac{\partial}{\partial k_l}\,\hat{g}^{\,c}_{\text{f}}({\bf
k})\right]\nonumber\\[3pt]
&+& \sum_{c=1}^8\sum_{\bf k}\frac{5ik_l}{4k^2}\left[A_Q^{c\,\mbox{\normalsize
$\star$}}({\bf k})A_R^c({\bf k}) - A_R^{c\,\mbox{\normalsize $\star$}}({\bf
k})A_Q^c({\bf k})\right]\nonumber\\[3pt]
&-& (1-\gamma)\sum_{c=1}^3\sum_{\bf
k}\frac{16ik^3}{m^3}\left[\frac{\partial}{\partial
k_l}\,A_Q^{c\,\mbox{\normalsize $\star$}}({\bf k})A_Q^c({\bf k}) -
A_Q^{c\,\mbox{\normalsize $\star$}}({\bf k})\frac{\partial}{\partial
k_l}\,A_Q^c({\bf k})\right].
\end{eqnarray}
Using the commutations rules given by Eqs.~(\ref{eq:AAdagger}) and
(\ref{eq:BBdagger}), we find that
\begin{equation}
\delta A^{c\dagger}({\bf k}) = \left[\frac{i{\sf
m}_c\epsilon_{ln}k_n}{k^2}\,A^{c\dagger}({\bf k}) -
\omega_c(k)\frac{\partial}{\partial k_l}\,A^{c\dagger}({\bf
k})\right]\delta\beta_l
\label{eq:delAdagger}
\end{equation}
and
\begin{equation}
\delta B^{\,c\dagger}({\bf k}) = \left[-\frac{i\bar{\sf
m}_c\epsilon_{ln}k_n}{k^2}\,B^{\,c\dagger}({\bf k}) -
\bar{\omega}_c(k)\frac{\partial}{\partial k_l}\,B^{\,c\dagger}({\bf
k})\right]\delta\beta_l.
\label{eq:delBdagger}
\end{equation}

Equations~(\ref{eq:delAdagger}) and (\ref{eq:delBdagger}) show that all the
massive gauge boson states\,---\,the single excitation mode $A^{c\dagger}({\bf
k})|0\rangle$ in the $(c=1,2,3)$ sectors with the residual $SU(2)$ invariance,
and the two excitation modes $A^{c\dagger}({\bf k})|0\rangle$ and
$B^{\,c\dagger}({\bf k})|0\rangle$ in the $(c=4,\ldots,8)$ `broken' doublet and
singlet sectors\,---\,transform {\it without any mixing\/} with other modes.
The phase factors $[{\sf m}_c\epsilon_{ln}k_n/k^2]\delta\beta_{l}$ and
$-[\bar{\sf m}_c\epsilon_{ln}k_n/k^2]\delta\beta_{l}$ generated by the boost
operator $K_l$, which appear in Eqs.~(\ref{eq:delAdagger}) and
(\ref{eq:delBdagger}), are the cocycles mentioned in Ref.~\cite{HLL3}. These
phase factors have no physical implications. The physically observable
consequence of Eqs.~(\ref{eq:delAdagger}) and (\ref{eq:delBdagger}) is that,
under a Lorentz transformation, the topologically massive gauge excitations
behave like the massive excitations of a scalar field\,---\,each topologically
massive gauge excitation transforms only into itself at a new space-time point.

\section{Conclusion}
In this paper we have presented a detailed analysis of the canonical
quantization of spontaneously broken topologically massive gauge theory. In
$2+1$ dimensions the possibility of including a Chern-Simons term in the gauge
field Lagrangian leads to new forms of mass-generating effects
for gauge fields. The resulting Chern-Simons-Higgs mechanism differs in
interesting ways from the conventional Higgs-Kibble mechanism, and in this
paper we have explored the Chern-Simons-Higgs mechanism by concentrating on
the relation between the quantized fields and their particle excitation
modes. We have found, by a series of unitary transformations, a consistent
particle-mode representation of the operator-valued fields and we have
constructed the corresponding Fock space which permits a consistent
interpretation of the diagonalized noninteracting Hamiltonian $H_0$ as an
energy operator for a system of noninteracting particles in a new vacuum
state. Within this Fock space, $H_0$ acts unitarily as a time translation
generator, in such a way that Gauss's law and the gauge condition are
manifestly preserved. We have computed the gauge field propagators as vacuum
expectation values of time-ordered products of the gauge field operators, and
formulated the corresponding perturbative expansion of the interacting
theory. We have chosen to present our analysis for a non-Abelian Chern-Simons
theory in which the original non-Abelian symmetry is spontaneously broken,
but with a residual non-Abelian symmetry in the broken vacuum. Such a
non-Abelian model clearly illustrates the interplay of the space-time and
algebraic features of the Chern-Simons-Higgs mechanism. This particular model
is also motivated by the question of its quantum consistency. Indeed, the
result reported in \cite{CDHL}, that the bare quantum consistency condition of
Deser-Jackiw-Templeton \cite{sdrj} is maintained at one-loop in such a broken
vacuum, was in fact first obtained by us using the techniques and
formalism described in this paper. An interesting further application would be
to the analysis of the non-Abelian versions of the self-dual Chern-Simons-Higgs
systems considered in \cite{klee}.

\acknowledgements
This research was supported by the Department of Energy under Grant
No.~DE-FG02-92ER40716.00.

\appendix
\section{Interaction lagrangian}
In this Appendix, we record the explicit expansions of the interaction
Lagrangians ${\cal L}_1$ and ${\cal L}_2$ in terms of real fields. These
interaction Lagrangians define the vertices required for perturbative
computations. When the Lagrangians given in Eqs.~(\ref{eq:calL1}) and
(\ref{eq:calL2}) are expanded in terms of the real fields in
Eq.~(\ref{eq:phiprime}) and the symmetry breaking mass scales in
Eq.~(\ref{eq:mass}), we obtain the following: the ${\cal O}(e)$ interaction
Lagrangian becomes
\begin{eqnarray}
{\cal L}_{1} &=& ef^{abc}F_{\mu\nu}^{a}A^{b\mu}A^{c\nu}
-{\frac{1}{3}}\,em\epsilon^{\mu\nu\rho}f^{abc}A_{\mu}^{a}
A_{\nu}^{b}A_{\rho}^{c} +
\displaystyle\sum_{a=4}^{7} eM_{\rm D}A^{a\mu}A_{\mu}^{a}\psi\nonumber\\
&+& \frac{2e}{\sqrt{3}}M_{\rm S}A^{8\mu}A_{\mu}^{8}\psi
+eM_{\rm D}d^{ab4}A^{a\mu}A_{\mu}^{b}\xi^{5}
-eM_{\rm D}d^{ab5}A^{a\mu}A_{\mu}^{b}\xi^{4}\nonumber\\
&+&eM_{\rm D}d^{ab6}A^{a\mu}A_{\mu}^{b}\xi^{7} - eM_{\rm
D}d^{ab7}A^{a\mu}A_{\mu}^{b}\xi^{6}\nonumber \\
&-&ie\left[\Psi_{1}^{\dagger}{\bf A}^{\mu}\cdot\mbox{\boldmath
$\tau$}\partial_{\mu}
\Psi_{1}-(\partial_{\mu}\Psi_{1})^{\dagger}{\bf A}^{\mu}\cdot\mbox{\boldmath
$\tau$}\Psi_{1}\right] \nonumber \\
&-&  ie\left[\Psi_{2}^{\dagger}(A^{4\mu}\tau^{1}
+A^{5\mu}\tau^{2})\partial_{\mu}\Psi_{2}-
(\partial_{\mu}\Psi_{2})^{\dagger}(A^{4\mu}\tau^{1}+
A^{5\mu}\tau^{2})\Psi_{2}\right] \nonumber\\
&-& ie\left[\Psi_{3}^{\dagger}(A^{6\mu}\tau^{1}
+A^{7\mu}\tau^{2})\partial_{\mu}\Psi_{3}-
(\partial_{\mu}\Psi_{3})^{\dagger}(A^{6\mu}\tau^{1}+
A^{7\mu}\tau^{2})\Psi_{3}\right]  \nonumber\\
& - &ie(\Phi^{\prime\dagger}A^{8\mu}\lambda^{8}\partial_{\mu}\Phi^{\prime}-
\partial_{\mu}\Phi^{\prime\dagger}A^{8\mu}\lambda^{8}\Phi^{\prime})\nonumber \\
 & - & \frac{e\mu^{2}}{M_{\rm D}}\,\psi\left[(\xi^{4})^{2}+(\xi^{5})^{2}+
(\xi^{6})^{2}+(\xi^{7})^{2}+(\xi^{8})^{2}+\psi^{2}\right]\nonumber\\
&+&2ief^{abc}A_{\mu}^{a}\sigma_{\text{f}}^{b}\partial^{\mu}
\sigma_{\text{p}}^{c}\label{eq:Lagone}
\end{eqnarray}
and the ${\cal O}(e^2)$ interaction Lagrangian becomes
\begin{eqnarray}
{\cal L}_{2} & = & -e^{2}f^{abc}f^{ade}A_{\mu}^{b}A^{d\mu}A_{\nu}^{c}
A^{e\nu}\nonumber\\
&+&{\frac{1}{3}}\,e^{2}A_{\mu}^{a}A^{a\mu}\left[(\xi^{4})^{2}+(\xi^{5})^{2}+
(\xi^{6})^{2}+(\xi^{7})^{2}+(\xi^{8})^{2}+\psi^{2}\right] \nonumber  \\
 & + & e^{2}d^{ab1}A^{a\mu}A_{\mu}^{b}(\xi^{5}\xi^{7}+\xi^{4}\xi^{6})
+ e^{2}d^{ab2}A^{a\mu}A_{\mu}^{b}(\xi^{5}\xi^{6}-\xi^{4}\xi^{7}) \nonumber \\
 & + &{\frac{1}{2}}\,e^{2}d^{ab3}A^{a\mu}A_{\mu}^{b}\left[(\xi^{4})^{2}
+(\xi^{5})^{2}-(\xi^{6})^{2}-(\xi^{7})^{2}\right] \nonumber \\
 & + & e^{2}d^{ab4}A^{a\mu}A_{\mu}^{b}(\xi^{5}\psi-\xi^{4}\xi^{8})
- e^{2}d^{ab5}A^{a\mu}A_{\mu}^{b}(\xi^{5}\xi^{8}+\xi^{4}\psi)  \nonumber\\
 & + & e^{2}d^{ab6}A^{a\mu}A_{\mu}^{b}(\xi^{7}\psi-\xi^{6}\xi^{8})
- e^{2}d^{ab7}A^{a\mu}A_{\mu}^{b}(\xi^{7}\xi^{8}+\xi^{6}\psi) \nonumber \\
 & + &{\frac{1}{2\sqrt{3}}}\,e^{2}d^{ab8}A^{a\mu}A_{\mu}^{b}
\left[(\xi^{4})^{2}+(\xi^{5})^{2}+
(\xi^{6})^{2}+(\xi^{7})^{2}-2(\xi^{8})^{2}-2\psi^{2}\right]  \nonumber \\
 &-& \frac{e^{2}\mu^{2}}{4\,M_{\rm D}^{2}}\left[(\xi^{4})^{2}+(\xi^{5})^{2}+
(\xi^{6})^{2}+(\xi^{7})^{2}+(\xi^{8})^{2}+\psi^{2}\right] ^{2}.
\label{eq:Lagtwo}
\end{eqnarray}
In these expressions, \mbox{\boldmath $\tau$} designates the Pauli spin
matrices, and ${\bf A}^{\mu}$ denotes the gauge field triplet $A^{a\mu}$
$(a=1,2,3)$ in the unbroken $SU(2)$ ``isospin'' subgroup.  The isospinors
$\Psi_a$ $(a=1,2,3)$ are the combinations of the Higgs field $\psi$ and the
$\xi^a$ fields given by Eqs.~(\ref{eq:psi1cap})--(\ref{eq:psi3cap}).

\end{document}